\documentclass[12pt, draftclsnofoot, onecolumn]{IEEEtran}
\usepackage{graphicx}
\usepackage{color}
\usepackage{graphicx}
\usepackage{epsf}
\usepackage{amsmath}
\usepackage{amsfonts}
\usepackage{latexsym}
\usepackage{subfigure}
\usepackage{multirow}
\usepackage{multicol}
\usepackage{amssymb}
\usepackage{cite}
\usepackage{float}
\usepackage{algorithm}
\usepackage{algpseudocode}
\usepackage{lipsum}

\allowdisplaybreaks
\title{{ Rate-Energy Region in Wireless Information and Power Transfer: New Receiver Architecture and Practical Modulation}
\date{\today}}
\author{\IEEEauthorblockN{Young-bin Kim, Dae Kyu Shin, \textit{Member, IEEE}, and Wan Choi, \textit{Senior Member, IEEE}}
\thanks{W. Choi is with School of Electrical Engineering, Korea Advanced Institute of Science and Technology (KAIST), Daejeon 34141, Korea (e-mail: wchoi@kaist.edu).}
\thanks{ Y.-b. Kim and D. K. Shin were with School of Electrical Engineering, KAIST, and are now with KDDI Research, Inc., Saitama 356-8502, Japan and with Samsung Electronics Co., Ltd, Suwon 16677, Korea, respectively. }}
\begin{document}
\maketitle

\vspace{-1in}
\begin{abstract}
When simultaneous wireless information and power transfer is carried out, a fundamental tradeoff between achievable rate and harvested energy exists because the received power is used for two different purposes. The tradeoff is well characterized by the rate-energy region, and several techniques have been proposed to improve the achievable rate-energy region. However, the existing techniques still have a considerable loss in either energy or rate and thus the known  achievable rate-energy regions are far from the ideal one. Deriving tight upper and lower bounds on the rate-energy region of our proposed scheme, we prove that the rate-energy region can be expanded almost to the ideal upper bound. Contrary to the existing techniques, in the proposed scheme, the information decoding circuit not only extracts amplitude and phase information but also combines the extracted information with the amplitude information obtained from the rectified signal. Consequently, the required energy for decoding can be minimized, and thus the proposed scheme achieves a near-optimal rate-energy region, which implies that the fundamental tradeoff in the achievable rate-energy region is nearly eliminated. To practically account for the theoretically achievable rate-energy region, we also present practical examples with an $M$-ary multi-level circular QAM with Gaussian maximum likelihood detection.
\end{abstract}


\section{Introduction} \label{sec:intro}
Energy efficient transmission is one of key considerations in recent wireless networks, such as wireless sensor networks, due to a limited lifetime of fixed energy supplies, e.g., batteries. In parallel, high costs and difficulty of frequent battery replacing motivates remote energy recharging technologies. Remote energy charing entials wireless power transfer (WPT)-enabled communications where wireless information transfer is combined with WPT. The WPT-enabled communication is in general classified into categories: simultaneous wireless information and power transfer (SWIPT) where energy harvesting and information decoding are simultaneously carried out at the receiver,  and wireless powered communication networks (WPCN) where wireless information is transmitted with the harvested energy.

SWIPT has been studied as a unified approach to energy harvesting and information decoding  \cite{ISIT_V2008, ISIT_GS2010}. In SWIPT, generally, there exists a fundamental tradeoff between achievable rate and harvested energy, which is characterized by the rate-energy region.
With a constraint on the amplitude of the transmit signal, finding the rate-energy region is known to be non-trivial according to input distributions \cite{ISIT_V2008}, while with an average power constraint on the transmit signal, the achievable rate-energy region can be identified \cite{ISIT_GS2010}. To expand the achievable rate-energy region, several approaches in receiver design have been investigated  \cite{CM_BHZ2015,TWC_LZC2013, TC_LZC2013, TC_ZZH2013}. Typically, the rate-energy tradeoff is optimized by either power split or time division between battery charging and information decoding. 
However, even with either opportunistic switching between WPT and information transfer in a time-division manner \cite{TWC_LZC2013} or partial energy utilization in both WPT and information transfer with an optimized  power split \cite{TWC_ZH2013, TC_LZC2013}, there exist fundamental limitations of simultaneous efficiency improvements in terms of both achievable rate and harvested energy  because the received energy is split for different purposes. In view of the rate-energy region, the amount of harvested energy with a time switching (TS) receiver or a power split (PS) receiver decreases as data rate increases.
  To minimize inefficiency resulted from the energy split for different purposes, an integrated information and energy (IIE) receiver was proposed in \cite{TC_ZZH2013}. In the IIE receiver, the received signal is rectified for charging battery and only a small portion of the rectified signal is used for decoding information. That is, in the IIE receiver, the amount of charged energy corresponds to amplitude information for information transfer. The IIE receiver  offers maximum capability of energy harvesting for some non-zero data rate, but there still exists a critical rate loss 
because information has to be carried over rectified signals. In view of the rate-energy region, an IIE receiver can harvest the maximum amount of energy if a data rate is below a certain threshold. However, it cannot harvest any energy if it tries to transmit information with data rate greater than the threshold, so the rate-energy region achieved by the IIE receiver is far from the optimal bound.
Recently, when multiple antennas are used at the IIE receiver, information decoding with rectified signals  was studied in \cite{JSEC_ZYH2015}. 
However, the conventional schemes for SWIPT still suffer form 
considerable energy and data rate losses, and thus the achievable rate-energy regions are still far from the ideal one.

In WPCN, wireless devices are first powered by WPT and then use the harvested energy to transmit their signals. Since energy harvesting was introduced in \cite{TWC_SM2010},  energy consumption strategies with the harvested energy have been studied in various 
communication scenarios, such as a point-to-point channel \cite{TC_HO2012, TC_YU2012, TWC_TY2012, TIT_OU2012}, a multiple access channel (MAC) \cite{JCAN_YS2012}, a broadcasting channel (BC) \cite{TC_OY2012}, a relay channel \cite{JSEC_HZ2013}, and an interference channel \cite{TC_SCK2015}. In these papers, wired energy supply from energy sources with restricted and irregular energy arrivals, \emph{i.e.,} solar, wind, etc., was assumed.
Contrary to the restricted and irregular energy sources, there have been extensive studies which use electromagnetic (EM) waves and radio-frequency (RF) signals for remote energy supply in various scenarios; inductive coupling \cite{Arxiv_L2011} and magnetic resonance coupling \cite{Science_K2007, ECCE_M2010} for near-field WPT, and RF energy transfer for far-field WPT \cite{TWC_ZH2013,CM_BHZ2015}. Joint resource allocation for WPT in a BC and information transmission in a MAC was optimized in \cite{WPCN1,WPCN2}. Furthermore, WPCNs with user cooperation~\cite{TC_GOYU2013,WPCN3,WPCN6}, full-duplex~\cite{WPCN4,WPCN5}, massive multiple-input multiple-output (MIMO)~\cite{WPCN8,WPCN9}, and cognitive techniques~\cite{WPCN11} were studied. 

The limited capability of the conventional SWIPT receivers is mainly due to the separated design of energy harvesting and information decoding, without sufficient considerations of interactions between them. This observation strongly motivates joint design of energy harvesting and information decoding  by taking account of the interplay between them.
 On the same line, 
the authors in \cite{varshney} argued that there is no thermodynamic limitation in achieving the ideal rate-energy region with power splitting, from examples of thermodynamically reversible computational devices.

In this context, we explore a new SWIPT receiver architecture to improve the efficiency of both WPT and information delivery. In particular, this paper proves that the rate-energy region can be considerably expanded almost to the ideal upper bound by the proposed receiver. While the rate-energy region achieved by the conventional SWIPT receiver was known to be far from the ideal upper bound, the derived
tight upper and lower bounds on the achievable rate-energy region of the proposed receiver demonstrate that the new achievable rate-energy region is significantly expanded compared to those of the conventional SWIPT receivers. 
Contrary to the IIE receiver, the proposed receiver exploits amplitude as well as phase for information transfer; the information decoding circuit extracts amplitude and phase information and combines the extracted information with the amplitude information obtained from the rectified signal. Because the amplitude information is partially obtained from the energy harvesting circuit and thus the  energy required for information decoding at the decoding circuit can be minimized. Consequently, the proposed scheme achieves near-optimal rate-energy region. That is, the fundamental tradeoff between WPT and information transfer in the achievable rate-energy region can be nearly eliminated, and SWIPT without sacrificing each other becomes possible.  
To practically account for the theoretically achievable rate-energy region, we also present practical examples of the rate-energy region improvement based on an $M$-ary multi-level circular QAM with multi-dimensional Gaussian maximum likelihood (ML) detection.  The proposed receiver structure is leveraged by signal constellations with multiple amplitude levels and different phases on each amplitude level. However, since taking account of all possible such constellations is impossible, we consider and optimize a structured one, multi-level circular QAM, as an example of such signal constellations. 

The rest of this paper is organized as follows.
In Section \ref{sec:system}, the system model for SWIPT  and the proposed receiver for improving rate-energy region are described. In Section \ref{sec:propose}, we analyze the rate-energy region achievable by the proposed receiver.  Practical examples of the rate-energy region improvement are by the proposed receiver presented  in Section \ref{sec:Constell}.
Finally, conclusions are drawn in Section \ref{sec:Con}.
\section{System model and an Unified Receiver Architecture for SWIPT} 
In this section, after describing the system model,
we propose a receiver architecture which integrates energy harvesting and information decoding while minimizing information and energy losses.

\label{sec:system}
\subsection{System Model}
\begin{figure}[t]
    \centerline{\includegraphics[width={0.8\columnwidth},height={0.17\columnwidth}]{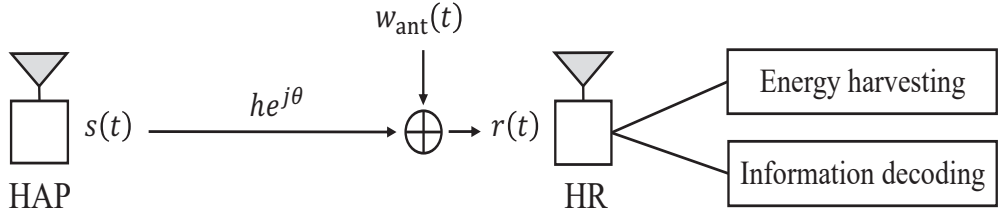}}
    \caption{A SWIPT system model}
    \label{fg:model}
\end{figure}
We consider a SWIPT system constituted by a hybrid access point (AP) and a hybrid receiver as shown in Fig. \ref{fg:model}. Both of the hybrid AP and the receiver have a single antenna each.
The symbol duration is $T$ and the corresponding signal bandwidth is assumed to $B = 1/T$ Hz. Let the complex baseband signal  transmitted from the hybrid AP be $x(t)=x_a(t)e^{jx_\phi(t)}$ where $\mathbb{E}\left[|x(t)|^2\right]=1$ and $x_a(t)$ and $x_\phi(t)$ denote amplitude and phase of $x(t)$, respectively.
Then, if the carrier frequency $f$ is much larger than the bandwidth, i.e., $f\gg B$, the passband signal transmitted from the hybrid AP becomes
\begin{align}
s(t)=\sqrt{2P}\mathfrak{R}\left\{x(t)e^{j2\pi ft}\right\}=\sqrt{2P}x_a(t)\cos\left(2\pi ft + x_\phi(t)\right)
\end{align} where the transmit signal is subject to an average power constraint given by $\mathbb{E}\left[|s(t)|^2\right]=P$.
Assuming an additive white Gaussian noise (AWGN) channel with a time invariant channel gain, the channel output is
\begin{align}
r(t)&= \sqrt{2}\mathfrak{R}\left\{y(t)\right\} = \sqrt{2}\mathfrak{R}\left\{\sqrt{P}he^{j\theta} x(t)e^{j2\pi ft} + z_{\mathrm{ant}}(t)e^{j2\pi ft}\right\} \\
&=\sqrt{2P}hx_a(t)\cos{\left(2\pi ft + x_{\phi}(t) + \theta\right)} + w_{\mathrm{ant}}(t)
\end{align}
where $h$ is a constant channel coefficient and $\theta \in [0, 2\pi)$ is a phase shift, $z_{\mathrm{ant}}(t)\sim\mathcal{CN}\left(0,\sigma_{\mathrm{ant}}^2\right)$ is a circular symmetric complex Gaussian noise, and $w_{\mathrm{ant}}(t)=\sqrt{2}\mathfrak{R}\left\{z_{\mathrm{ant}}(t)e^{j2\pi ft}\right\}$ is the corresponding passband Gaussian noise. The one-sided noise power spectral density is defined as $N_0=\frac{\sigma_{\mathrm{ant}}^2}{B}$. Our channel model with a constant channel coefficient corresponds to a frequency non-selective static or quasi-static channel, which typically occurs with narrow band signals in low mobility environments. The analysis with this channel model builds an analytic framework to obtain the ergodic rate-energy region in frequency non-selective fast fading channels. It might also be applicable and extended to frequency selective channels since each subcarrier experiences a frequency non-selective channel if orthogonal frequency division multiplexing (OFDM) is adopted. 
 
A hybrid receiver consists of two parts: information decoding and energy harvesting which are described in detail below.


\subsubsection{Information Decoding} \label{sec:ID}

\begin{figure}[t]
\centering
\begin{subfigure}[][Information decoding]
    {\includegraphics[width={0.84\columnwidth},height={0.3\columnwidth}]{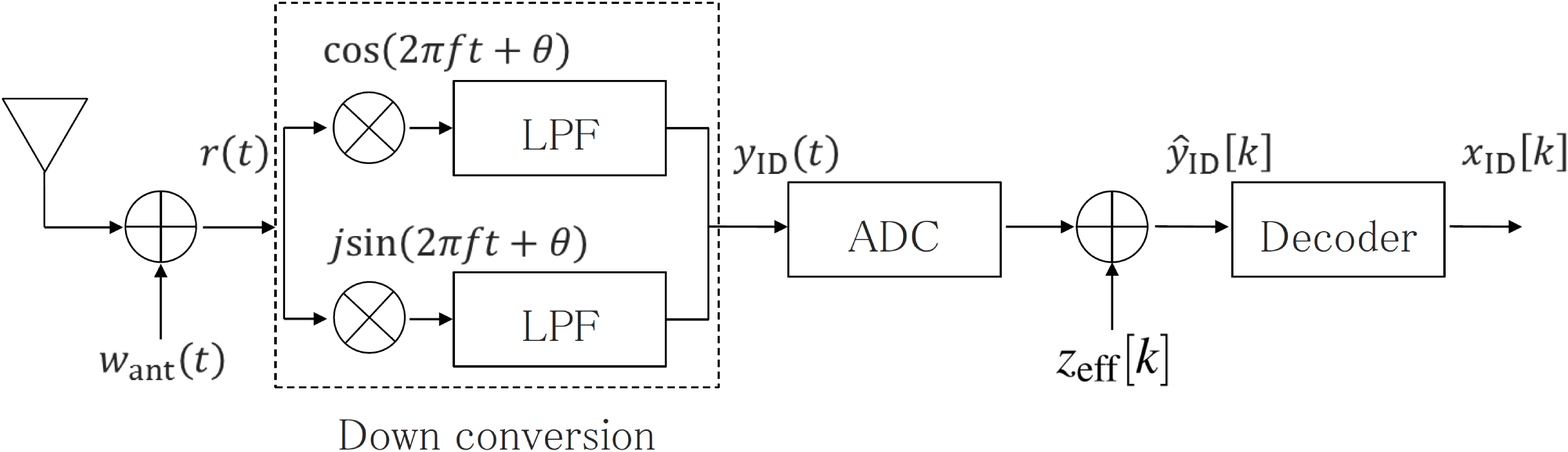}}
\end{subfigure}
\begin{subfigure}[][Energy harvesting]
    {\includegraphics[width={0.79\columnwidth},height={0.17\columnwidth}]{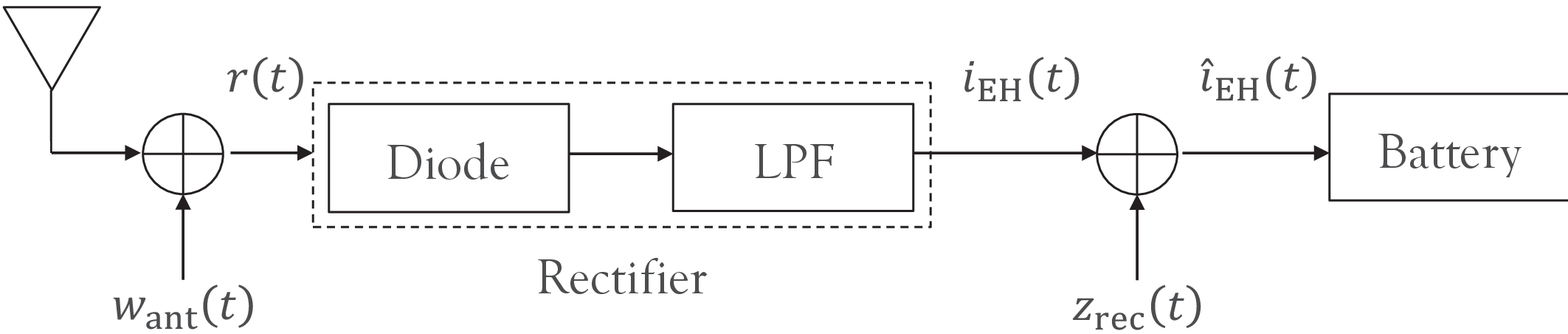}}
\end{subfigure}
    \caption{Typical signal processing for information decoding and energy harvesting at a hybrid receiver.}
    \label{fg:Rcv_IDEH}
\end{figure}

Fig. \ref{fg:Rcv_IDEH}(a) depicts optimal signal processing for information decoding; the received signal is first converted down to the baseband signal and then quantized by analog-to-digital conversion (ADC). Assuming quantization errors follow Gaussian distribution, the quantization error and additional noise signals at down-converter and ADC can be modeled together as  a circularly symmetric noise $z_{\mathrm{eff}}[k] \sim \mathcal{CN}\left(0,\sigma_{\mathrm{eff}}^2\right)$.
Consequently, the equivalent baseband signal of the decoder input  at time index $k$ is
$\hat{y}_{\mathrm{ID}}[k]= x_{\mathrm{ID}}[k] + z_{\mathrm{ID}}[k]$, 
where $z_{\mathrm{ID}}[k] \sim \mathcal{CN}\left(0,\sigma_{\mathrm{ant}}^2 + \sigma_{\mathrm{eff}}^2\right)$ and $x_{\mathrm{ID}}[k]$ and $y_{\mathrm{ID}}[k]$ denote the channel input and output at the information decoder, respectively. 
\subsubsection{Energy Harvesting} \label{sec:EH}
Fig. \ref{fg:Rcv_IDEH}(b) exhibits optimal signal processing for energy harvesting. Contrary to the receiver for information decoding, the RF band signal is rectified to obtain the direct current (DC) signal and can be built with a Shottky diode and a passive low-pass filter (LPF) as in \cite{TC_ZZH2013}. After passing the Shottky diode, the output current becomes
$i_{\mathrm{EH}}(t)=I_s\left(e^{\gamma r(t)} - 1\right) \approx c_1r(t) + c_2r^2(t)$
where $I_s$ is the saturation current; $\gamma$ is the reciprocal of the thermal voltage of the diode; $c_k=\frac{I_s\gamma^k}{k!}$, $k\in\mathbb{Z}_{+}$, which is given from the Taylor series expansion of the exponential function.
The approximation is tight because $\gamma r(t)$ is assumed to be close to zero in general.
Then,
after LPF which removes  high frequency components of the signal centered at $f$ and $2f$, the rectified signal is obtained as
\begin{align} \label{eq:output_rec}
\hat{i}_{\mathrm{EH}}(t) = c_2\mu_{\mathrm{EH}}^2(t) + z_{\mathrm{rec}}(t) = c_2\mu_{\mathrm{EH,I}}^2(t) + c_2\mu_{\mathrm{EH,Q}}^2(t) + z_{\mathrm{rec}}(t),
\end{align} 
where $\mu_{\mathrm{EH}}(t)=\sqrt{\mu_{\mathrm{EH,I}}^2(t) + \mu_{\mathrm{EH,Q}}^2(t)}$ with
$\newline\mu_{\mathrm{EH,I}}(t)=\sqrt{P}h x_a(t)\cos\left(x_\phi(t) + \theta\right) + z_{\mathrm{I}}(t)$ and
$\mu_{\mathrm{EH,Q}}(t)=\sqrt{P}h x_a(t)\sin\left(x_\phi(t) + \theta\right) + z_{\mathrm{Q}}(t)$, and $z_{\mathrm{rec}}(t) \sim \mathcal{N}\left(0,\sigma_{\mathrm{rec}}^2\right)$ denotes the additional noise at the rectifier.
 $z_{\mathrm{I}}(t) \sim \mathcal{N}\left(0,\sigma_{\mathrm{ant}}^2/2\right)$ and $z_{\mathrm{Q}}(t) \sim \mathcal{N}\left(0,\sigma_{\mathrm{ant}}^2/2\right)$ denote the in-phase and quadrature components of the complex baseband antenna noise $z_{\mathrm{ant}}(t)$, respectively.

 
Since $c_2$ is a constant specified by the diode, we assume that $c_2=1$ for convenience as \cite{TC_ZZH2013}. We also assume the amount of harvested energy from noise is negligible since it is relatively marginal and the length of symbol period is one. If the
 whole received signal is used for energy harvesting under the assumptions, the amount of energy charged at battery is given by
$Q_{\mathrm{EH}}=\zeta\mathbb{E}\left[\hat{i}_{\mathrm{EH}}(t)\right] = \zeta h^2P~(J)$
where $\zeta$ is a DC signal to energy conversion efficiency by practical limitations in saving energy, $\zeta \in \left(0,1\right]$. Note that power and energy can be interchangeable throughout the paper under the assumption that the length of symbol period is unit.
\subsection{Preliminary: Ideal Outer Bound on the Rate-Energy Region} \label{sec:region}
The outer bound of achievable rate-energy region is defined as
\begin{align} \label{eq:region_def}
\mathcal{C}_{\mathrm{R-E}}^{outer}(P) = \left\{\left(R,Q\right)\bigg|R\leq \log_2\left(1+\frac{h^2P}{\sigma_{\mathrm{ant}}^2 + \sigma_{\mathrm{eff}}^2}\right),~Q\leq \zeta h^2P\right\}.
\end{align}
Because the whole received signal cannot be used for one purpose only in SWIPT,
the rate-energy region practically achievable has been known to be much smaller than the outer bound in \eqref{eq:region_def}. The objective of our paper is to expand the achievable rate-energy region to be close to the ideal upper bound in \eqref{eq:region_def}.

\subsection{Proposed Receiver Architecture} \label{sec:propose_sub1}

\begin{figure}[t]
    \centerline{\includegraphics[width={0.9\columnwidth},height={0.36\columnwidth}]{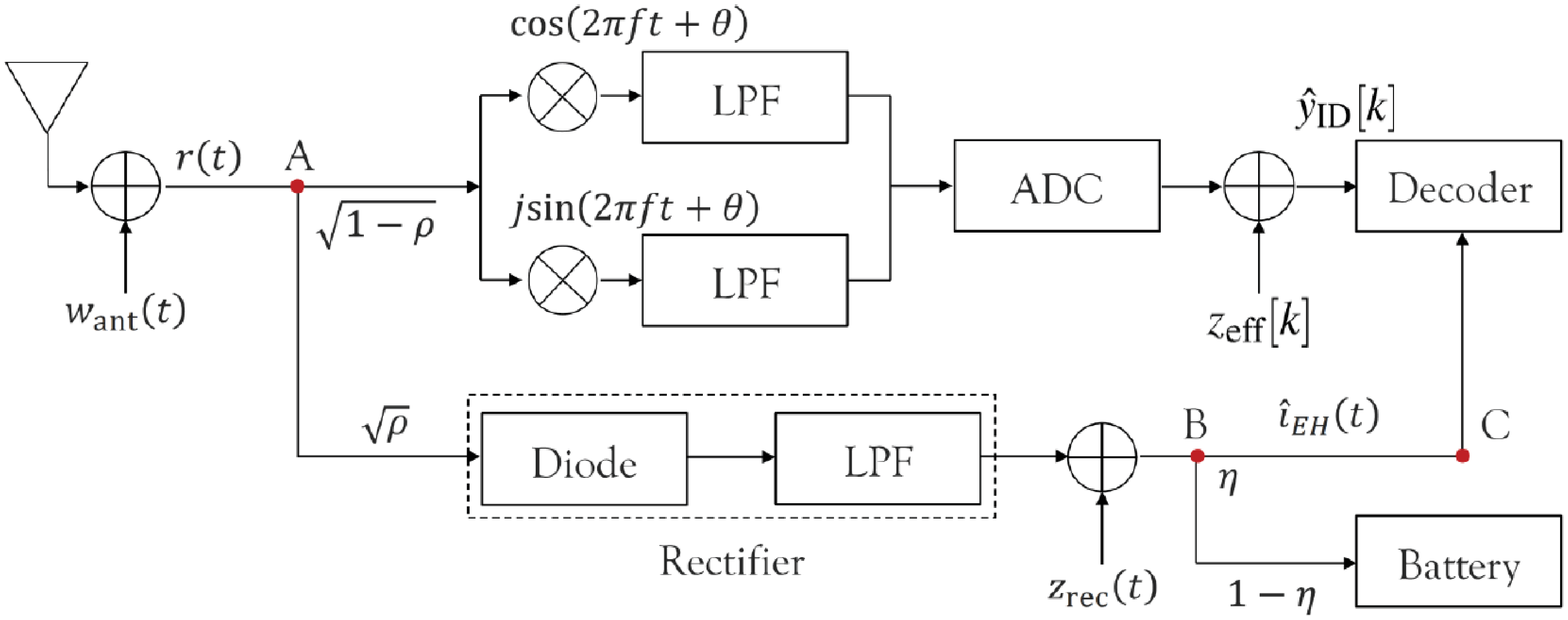}}
    \caption{The proposed receiver architecture}
    \label{fg:Rcv_pro}
\end{figure}

As shown in  Fig. \ref{fg:Rcv_pro}, the proposed receiver architecture consists of two signal processing paths as the conventional power split architecture, but the rectified signal is exploited not only for energy harvesting but also for decoding amplitude information. On the other hand, the baseband signal processing part extracts amplitude and phase information, and combines
the extracted information with the information obtained from the rectified signal. Specifically, the received signal $r(t)$ is split at point A according to the power split portions  $1-\rho$ and $\rho$. The signal with the power portion $\rho$ is rectified and then is again split into two paths at point B according to portions $1-\eta$ and $\eta$; one for battery recharging and the other for extracting amplitude information for the decoder.
In order to transfer information over phase as well as amplitude, the receiver has the conventional baseband signal processing path. At the decoder, the information obtained from baseband signal processing is combined with the amplitude information acquired from the rectified signal. In this way, the information and energy losses can be reduced; the power portion $1-\rho$ can be reduced without decreasing the achievable rate because the amplitude information can be obtained from both of the baseband and  rectified signals while the baseband signal processing part can focus on decoding phase information. 


\section{Rate-Energy Region Analysis} \label{sec:propose}

In this section, we derive and show the rate-energy region achieved by the proposed architecture nearly achieves the rate-energy outer bound in \eqref{eq:region_def}.

\subsection{Achievable Rate-Energy Region}
From the output signal of the rectifier given in \eqref{eq:output_rec}, the signal at point C in the proposed receiver is represented as
\begin{align} \label{eq:decoder1}
\hat{i}_{\mathrm{EH}}(t) &= \eta \cdot\left( \left| \sqrt{\rho h^2P}x_a(t)e^{j\left(\theta+x_\phi(t)\right)} + \sqrt{\rho}z_{\mathrm{ant}}(t) \right|^2 + z_{\mathrm{rec}}(t) \right) \nonumber\\
&= \eta\cdot\left( \left|\sqrt{\rho h^2P}x_a(t) + \sqrt{\rho}z_{\mathrm{ant}}(t)\right|^2 + z_{\mathrm{rec}}(t) \right).
\end{align}
Because SNR for $x_a(t)$ in $\hat{i}_{\mathrm{EH}}(t)$ does not changed for any $\eta ~(>0)$, an arbitrary small positive portion can be assumed to be used, i.e., $\eta\rightarrow0$.
Consequently, up to $\zeta\rho h^2P$ energy can be saved at the battery, i.e., $Q \approx \zeta\rho h^2P$.

For $n$ channel uses, the mutual information obtained with the proposed receiver is given by
\begin{align} \label{eq:info}
I\left( \hat{i}_{\mathrm{EH}}^n,\hat{y}_{\mathrm{ID}}^n;x_a^n,x_\phi^n\right)
&=I\left( \hat{i}_{\mathrm{EH}}^n;x_a^n,x_\phi^n\right)+I\left( \hat{y}_{\mathrm{ID}}^n;x_a^n,x_\phi^n|\hat{i}_{\mathrm{EH}}^n\right)\nonumber\\
&=\underbrace{I\left( \hat{i}_{\mathrm{EH}}^n;x_a^n \right)}_{\mathrm{ from~the~rectified~signal} } + \underbrace{I\left( \hat{y}_{\mathrm{ID}}^n;x_a^n \vert \hat{i}_{\mathrm{EH}}^n \right) + I\left( \hat{y}_{\mathrm{ID}}^n; x_\phi^n \vert \hat{i}_{\mathrm{EH}}^n,x_a^n \right)}_{\mathrm{from~the~baseband~signal}}
\end{align}
where $f^n=\left\{f(1),\dots,f(n)\right\}$ and
\eqref{eq:info} comes from $I\left( \hat{i}_{\mathrm{EH}}^n;x_\phi^n|x_a^n \right)=0$. 
The first term and the last two terms in \eqref{eq:info} represent mutual information from the rectified signal and the baseband signal at the proposed receiver, respectively.

\subsubsection{Outer Bound}
From Fano's inequality, the achievable rate from the rectified signal is upper bounded by \cite{TC_ZZH2013}
\begin{align} \label{rec_rate_upper}
nR_{\mathrm{EH}} \leq I\left( \hat{i}_{\mathrm{EH}}^n;x_a^n \right) + n\epsilon_n \leq
\left\{
\begin{array}{ll}
nC_{\mathrm{OIC}} + n\epsilon_n, & \sigma_{\mathrm{ant}}^2=0, \\
nC_{\mathrm{NAC}} + n\epsilon_n, & \sigma_{\mathrm{rec}}^2=0
\end{array}
\right.
\end{align}
where $C_{\mathrm{OIC}}$ is the capacity of the optimal intensity channel which corresponds to the case when the rectified signal is obtained without antenna noise, i.e., $z_{\mathrm{ant}}(t)=0$  in \eqref{eq:decoder1}; $C_{\mathrm{NAC}}$ is the capacity of the non-coherent AWGN channel which corresponds to the case without rectifier noise, i.e., $z_{\mathrm{rec}}(t)=0$  in \eqref{eq:decoder1}.

It is known in \cite{TIT_LMW2009} that $C_{\mathrm{OIC}}$ is bounded above by
\begin{align}
C_{\mathrm{OIC}} \leq& \log_2\left(\beta e^{-\frac{\delta^2}{2\sigma_{\mathrm{rec}}^2}} + \sqrt{2\pi}\sigma_{\mathrm{rec}}\mathcal{Q}\left(\frac{\delta}{\sigma_{\mathrm{rec}}}\right) \right) + \frac{1}{2}\mathcal{Q}\left(\frac{\delta}{\sigma_{\mathrm{rec}}}\right) + \frac{1}{\beta}\left( \delta + \rho h^2P + \frac{\sigma_{\mathrm{rec}}e^{-\frac{\delta^2}{2\sigma_{\mathrm{rec}}^2}} }{\sqrt{2\pi}} \right) \nonumber \\
& \quad+ \frac{\delta e^{-\frac{\delta^2}{2\sigma_{\mathrm{rec}}^2}}}{2\sqrt{2\pi}\sigma_{\mathrm{rec}}} + \frac{\delta^2}{2\sigma_{\mathrm{rec}}^2}\left\{ 1 - \mathcal{Q}\left(\frac{\delta + \rho h^2P}{\sigma_{\mathrm{rec}}}\right) \right\} - \frac{1}{2}\log_2{2\pi e\sigma_{\mathrm{rec}}^2} \label{eq:OIC}
\end{align}
where $\beta$ and $\delta$ are free parameters, $\beta>0$ and $\delta\geq0$, and $Q(x)=\frac{1}{\sqrt{2\pi}}\int_{x}^{\infty}e^{\frac{-\tau^2}{2}}d\tau$. The upper bound in \eqref{eq:OIC} becomes tight with parameters
\begin{align}
&\delta = \sigma_{\mathrm{rec}}\log_2\left( 1 + \frac{\rho h^2P}{ \sigma_{\mathrm{rec}} } \right), \label{eq:del} \\
&\beta = \frac{1}{2}\left( \delta + \rho h^2P + \frac{ \sigma_{\mathrm{rec}} }{\sqrt{2\pi}}e^{ -\frac{\delta^2}{ 2\sigma_{\mathrm{rec}}^2 } } \right) \nonumber\\
&~~ + \frac{1}{2}\left\{\! \left(\! \delta + \rho h^2P + \frac{ \sigma_{\mathrm{rec}} }{\sqrt{2\pi}}e^{ -\frac{\delta^2}{ 2\sigma_{\mathrm{rec}}^2 } } \right)^2 \!\! + \!4\!\left(\! \delta + \rho h^2P + \frac{ \sigma_{\mathrm{rec}} }{\sqrt{2\pi}}e^{ -\frac{\delta^2}{ 2\sigma_{\mathrm{rec}}^2 } } \!\right)\sqrt{2\pi}\sigma_{\mathrm{rec}}e^{ \frac{\delta^2}{ 2\sigma_{\mathrm{rec}}^2 } }\mathcal{Q}\left( \frac{\delta}{ \sigma_{\mathrm{rec}} } \right)\! \right\}^{\frac{1}{2}}, \label{eq:bet}
\end{align}
which ensure only a marginal difference from the lower bound of $C_{\mathrm{OIC}}$,  and the difference diminishes as the transmit power goes to infinity  \cite{TIT_LMW2009}.
Therefore, if we adopt the values of $\beta$ and $\delta$ in \eqref{eq:del} and \eqref{eq:bet}, $C_{\mathrm{OIC}}$ used in \eqref{rec_rate_upper} can be evaluated well in the proposed receiver architecture.

On the other hand, an upper bound of $C_{\mathrm{NAC}}$ can be obtained by maximizing the achievable rate over all possible input distributions and then is given by  \cite{TIT_KS2004}
\begin{align}
C_{\mathrm{NAC}} \leq \frac{1}{2}\log_2\left( 1 + \frac{\rho h^2P}{\sigma_{A}^2} \right) + \frac{1}{2}\left( \log_2{\frac{2\pi}{e}} - c_E\log_2e \right) \label{eq:NAC}
\end{align}
where $c_E=\int_1^{\infty} \left( \frac{1}{\lfloor \tau \rfloor} - \frac{1}{\tau} \right) d\tau$ is the Euler-Mascheroni constant.
The tightness of this upper bound \eqref{eq:NAC} is numerically presented in \cite{TIT_KS2004, C_L2002} for high SNR.
 The upper bound in \eqref{eq:NAC} shows less than 0.2 nats difference from the capacity $C_{\mathrm{NAC}}$ and becomes tighter as $P\rightarrow\infty$.

Consequently, from \eqref{rec_rate_upper} with $n\rightarrow\infty$, the error probability goes to zero and the achievable rate from the rectified signal $R_{\mathrm{EH}}$ in our proposed receiver is bounded above by
\begin{align} \label{bound1}
R_{\mathrm{EH}} \leq \min \big\{ \eqref{eq:OIC}, \eqref{eq:NAC} \big\}
\end{align}
with parameters $\beta$ and $\delta$ in \eqref{eq:del} and \eqref{eq:bet}. According to input distributions, we can find another upper bound on the achievable rate as
\begin{align}\label{bound2}
nR_{\mathrm{EH}}\leq I\left( \hat{i}_{\mathrm{EH}}^n;x_a^n \right) + n\epsilon_n \leq I\left( \sqrt{\rho}r^n;x_a^n \right) + n\epsilon_n \overset{(a)}{\leq} n\log_2\left(1+\frac{\rho h^2P}{\rho\sigma_{\mathrm{ant}}^2+\sigma_{\mathrm{rec}}^2}\right) + n\epsilon_n
\end{align}
where $(a)$ is the maximum achievable rate from information decoding with $\sqrt{\rho}r(t)$ under Gaussian signaling (i.e., (complex) Gaussian distributed input signals).

The information extracted from the rectified signal is passed to the decoder, and helps the decoder decode the transmitted message from the $\sqrt{1-\rho}$ portion of the received signal. As a result, the achievable rate is upper bounded as
\begin{align}
nR_{\mathrm{ID}} &\leq I\left( \hat{y}_{\mathrm{ID}}^n;x_a^n \vert \hat{i}_{\mathrm{EH}}^n \right) + I\left( \hat{y}_{\mathrm{ID}}^n;x_\phi^n \vert x_a^n \right) + n\epsilon_n \\
&\overset{(b)}{\leq}I\left( \hat{y}_{\mathrm{ID}}^n;x_a^n \right) + I\left( \hat{y}_{\mathrm{ID}}^n;x_\phi^n \vert x_a^n \right) + n\epsilon_n \label{eq:R_ID_b}\\
& = I\left( \hat{y}_{\mathrm{ID}}^n;x_a^n, x_\phi^n  \right) +  n\epsilon_n \\
&\overset{(c)}{\leq} n \log_2\left(1+\frac{\left(1-\rho\right)h^2P}{\left(1-\rho\right)\sigma_{\mathrm{ant}}^2+\sigma_{\mathrm{eff}}^2}\right) + n\epsilon_n\label{eq:R_ID_c}
\end{align}
where $(b)$ follows from $h\left( \hat{y}_{\mathrm{ID}}^n \vert \hat{i}_{\mathrm{EH}}^n \right) \leq  h\left( \hat{y}_{\mathrm{ID}}^n \right)$ and $h\left( \hat{y}_{\mathrm{ID}}^n \vert x_a^n,\hat{i}_{\mathrm{EH}}^n \right) = h\left( \hat{y}_{\mathrm{ID}}^n \vert x_a^n \right)$; the equality in $(c)$ holds with Gaussian distributed input signals.

Combining \eqref{bound1}, \eqref{bound2}, and \eqref{eq:R_ID_c} with $n\rightarrow\infty$, the achievable rate with the rectified signal and the baseband signal in the proposed receiver is bounded above by
\begin{align} \label{eq:infosum1}
R \leq \min\Bigg\{ \eqref{eq:OIC}, ~\eqref{eq:NAC}, ~\log_2\left(1+\frac{\rho h^2P}{\rho\sigma_{\mathrm{ant}}^2+\sigma_{\mathrm{rec}}^2}\right) \Bigg\} + \log_2\left(1+\frac{\left(1-\rho\right)h^2P}{\left(1-\rho\right)\sigma_{\mathrm{ant}}^2+\sigma_{\mathrm{eff}}^2}\right)
\end{align}
where $R=R_{\mathrm{EH}} + R_{\mathrm{ID}}$.

On the other hand, another upper bound of the achievable rate $R$ is derived from the data processing inequality as
\begin{IEEEeqnarray}{ll}\label{eq:infosum2}
\IEEEyesnumber\IEEEyessubnumber*
   nR   &\leq ~I\left( \hat{i}_{\mathrm{EH}}^n;x_a^n \right) + I\left( \hat{y}_{\mathrm{ID}}^n;x_a^n \vert \hat{i}_{\mathrm{EH}}^n \right) + I\left( \hat{y}_{\mathrm{ID}}^n;x_\phi^n \vert x_a^n \right)+n\epsilon_n \label{eq:infosum2_0}\\
    & =~I\left( \hat{i}_{\mathrm{EH}}^n, \hat{y}_{\mathrm{ID}}^n;x_a^n \right) + I\left( \hat{y}_{\mathrm{ID}}^n;x_\phi^n \vert x_a^n \right)+n\epsilon_n \label{eq:infosum2_1}\\
 &\overset{(d)}{\leq} ~I\left( \sqrt{\rho}r^n, \sqrt{1-\rho}r^n;x_a^n \right) + I\left( \sqrt{1-\rho}r^n;x_\phi^n \vert x_a^n \right)+n\epsilon_n \label{eq:infosum2_2}\\
 &= ~I\left( \sqrt{\rho}r^n;x_a^n \right) + I\left( \sqrt{1-\rho}r^n;x_a^n \vert \sqrt{\rho}r^n \right) + I\left( \sqrt{1-\rho}r^n;x_\phi^n \vert x_a^n \right)+n\epsilon_n \label{eq:infosum2_3}\\
    &\leq ~I\left( \sqrt{\rho}r^n;x_a^n \right) + I\left( \sqrt{1-\rho}r^n;x_a^n \vert \sqrt{\rho}r^n \right) + I\left( \sqrt{1-\rho}r^n;x_\phi^n \vert x_a^n \right) \IEEEnonumber\\
&\quad\quad + I\left( \sqrt{\rho}r^n;x_\phi^n \vert x_a^n, \sqrt{1-\rho}r^n \right)+n\epsilon_n \label{eq:infosum2_4}\\
 &= ~I\left( \sqrt{\rho}r^n,\sqrt{1-\rho}r^n;x_a^n,x_\phi^n \right)+n\epsilon_n \label{eq:infosum2_5}\\ &= ~I\left( r^n;x^n \right)+n\epsilon_n \label{eq:infosum2_6}\\
 &\overset{(e)}{\leq} ~n\log_2\left(1+\frac{h^2P}{\sigma_{\mathrm{ant}}^2+\sigma_{\mathrm{eff}}^2}\right)+n\epsilon_n \label{eq:infosum2_7}
\end{IEEEeqnarray}
where Markov chains $x_a^n,x_\phi^n\rightarrow r^n\rightarrow \sqrt{\rho}r^n \rightarrow \hat{i}_{\mathrm{EH}}^n$ and $x_a^n,x_\phi^n\rightarrow r^n\rightarrow \sqrt{1-\rho}r^n \rightarrow \hat{y}_{\mathrm{ID}}^n$ hold; $(d)$ is given from data processing inequality based on the Markov chains; $(e)$ holds with a Gaussian input distribution.

Therefore, from \eqref{eq:infosum1} and \eqref{eq:infosum2}, the maximum rate-energy region with the proposed receiver architecture is obtained as
\begin{align} \label{eq:region}
\mathcal{C}_{\mathrm{R-E}}(P) = \Bigg\{&\left(R,Q\right)\bigg|R\leq \min\left\{{C_{\mathrm{OIC}}},{C_{\mathrm{NAC}}},\log_2\left(1+\frac{\rho h^2P}{\rho\sigma_{\mathrm{ant}}^2+\sigma_{\mathrm{rec}}^2}\right)\right\} \nonumber\\
&+\log_2\left(1+\frac{\left(1-\rho\right)h^2P}{\left(1-\rho\right)\sigma_{\mathrm{ant}}^2+\sigma_{\mathrm{eff}}^2}\right),~R\leq \log_2\left(1+\frac{h^2P}{\sigma_{\mathrm{ant}}^2+\sigma_{\mathrm{eff}}^2}\right), \nonumber\\
&Q\leq \zeta\rho h^2P.\Bigg\}
\end{align}

\subsubsection{Inner Bound}
The achievable rate with the proposed receiver is certainly lower than the mutual information in
\eqref{eq:info}, which is maximized over all possible input distributions but is surely higher than or equal to that with a specific input distribution. Therefore, we can obtain a lower bound of the achievable rate with a specific distribution of the input  $x^n$ as
\begin{align}
&\max_{p(x)}\left\{I\left( \hat{i}_{\mathrm{EH}}^n;x_a^n \right) + I\left( \hat{y}_{\mathrm{ID}}^n;x_a^n \vert \hat{i}_{\mathrm{EH}}^n \right) + I\left( \hat{y}_{\mathrm{ID}}^n;x_\phi^n \vert x_a^n \right)\right\} + n\epsilon_n\\
&\geq \max_{p(x)}\big\{nR \big\}\\
&\geq I\left( \hat{i}_{\mathrm{EH}}^n;\overline{x}_a^n \right) + I\left( \hat{y}_{\mathrm{ID}}^n;\overline{x}_a^n \vert \hat{i}_{\mathrm{EH}}^n \right) + I\Big( \hat{y}_{\mathrm{ID}}^n;\overline{x}_\phi^n \vert \overline{x}_a^n \Big) + n\epsilon_n\label{eq:R_low}
\end{align}
where $p(x)$ is the distribution of $x^n$ and $\overline{x}_a^n$ and $\overline{x}_\phi^n$ are input variables with the specific distribution of $x^n$.

To obtain a specified lower bound of the achievable rate in \eqref{eq:R_low}, we consider a Gaussian distributed input $x^n$ as a specific distribution. Note that
since the last two terms  $I\left( \hat{y}_{\mathrm{ID}}^n;\overline{x}_a^n \vert \hat{i}_{\mathrm{EH}}^n \right) + I\left( \hat{y}_{\mathrm{ID}}^n;\overline{x}_\phi^n \vert \overline{x}_a^n \right)$ correspond to the achievable rate from baseband signal processing, they are well known to be maximized with the Gaussian input distribution and thus
become
\begin{align}
I\left( \hat{y}_{\mathrm{ID}}^n;\overline{x}_a^n \vert \hat{i}_{\mathrm{EH}}^n \right) + I\left( \hat{y}_{\mathrm{ID}}^n;\overline{x}_\phi^n \vert \overline{x}_a^n \right) = \log_2\left(1+\frac{\left(1-\rho\right)h^2P}{\left(1-\rho\right)\sigma_{\mathrm{ant}}^2+\sigma_{\mathrm{eff}}^2}\right)
\end{align} with the Gaussian input assumption as (\ref{eq:R_ID_c}).
On the other hand, note that the first term in \eqref{eq:R_low} $I\left( \hat{i}_{\mathrm{EH}}^n;\overline{x}_a^n \right)$ which denote the achievable rate from the rectified signal is not maximized with the Gaussion distributed input since Gaussian input distribution is not optimal in a mixed noisy channel with Chi-square noise $|\sqrt{\rho}z_{\mathrm{ant}}^n|^2$ and AWGN $z_{\mathrm{rec}}^n$. However, unfortunately, closed form of $I\left( \hat{i}_{\mathrm{EH}}^n;\overline{x}_a^n \right)$ with a Gaussian input distribution is not available. 

\begin{figure}[t]
    \centerline{\includegraphics[width={0.8\columnwidth},height={0.6\columnwidth}]{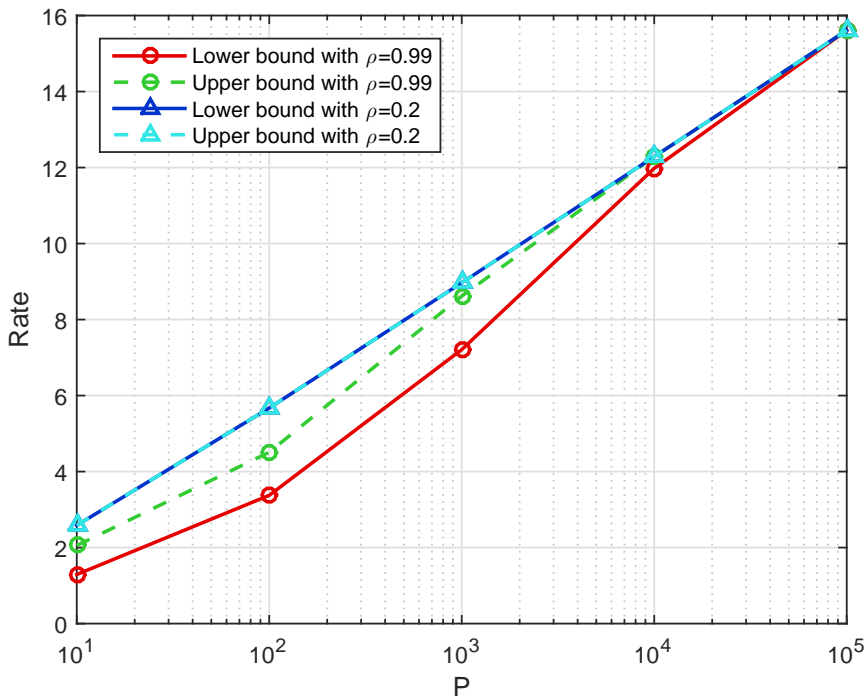}}
    \caption{Capacity bounds of the received signal in the unified receiver with $h=1$, $\sigma_{\mathrm{ant}}^2=\sigma_{\mathrm{rec}}^2=\sigma_{\mathrm{eff}}^2=1$, and $\zeta=0.6$ with Gaussian input distribution.}
    \label{fg:bound}
\end{figure}

The gap between the specified lower bound in \eqref{eq:R_low} with a Gaussian input distribution and the upper bound  in \eqref{eq:region} with suboptimal parameters $\beta$ and $\delta$ from \eqref{eq:del} and \eqref{eq:bet}, which minimize the upper bound of $C_{\mathrm{OIC}}$ in \eqref{eq:OIC}, diminishes as the transmit power increases as shown in Fig. \ref{fg:bound}. Note that 
the gap between the lower and upper bounds is determined  mainly by $I\left( \hat{i}_{\mathrm{EH}}^n;\overline{x}_a^n \right)$ with Gaussian input distribution.
 When the portion of the rectified signal is high, \emph{i.e.,} $\rho=0.99$, since the value of $I\left( \hat{i}_{\mathrm{EH}}^n;\overline{x}_a^n \right)$ becomes dominant, the gap between the upper and lower bounds in Fig. \ref{fg:bound} is large. On the contrary, when the portion of the rectified signal is relatively low, \emph{i.e.,} $\rho=0.2$, the value of $I\left( \hat{i}_{\mathrm{EH}}^n;\overline{x}_a^n \right)$  is marginal. Consequently, when $\rho=0.2$, the upper and lower bounds almost coincide with each other, which implies that the actual achievable rate can be represented as either the upper bound or the lower bound.  Moreover,
a proper input distribution instead of the Gaussian input distribution might be able to further reduce the gap.

\begin{figure}[t]
    \centerline{\includegraphics[width={0.8\columnwidth},height={0.6\columnwidth}]{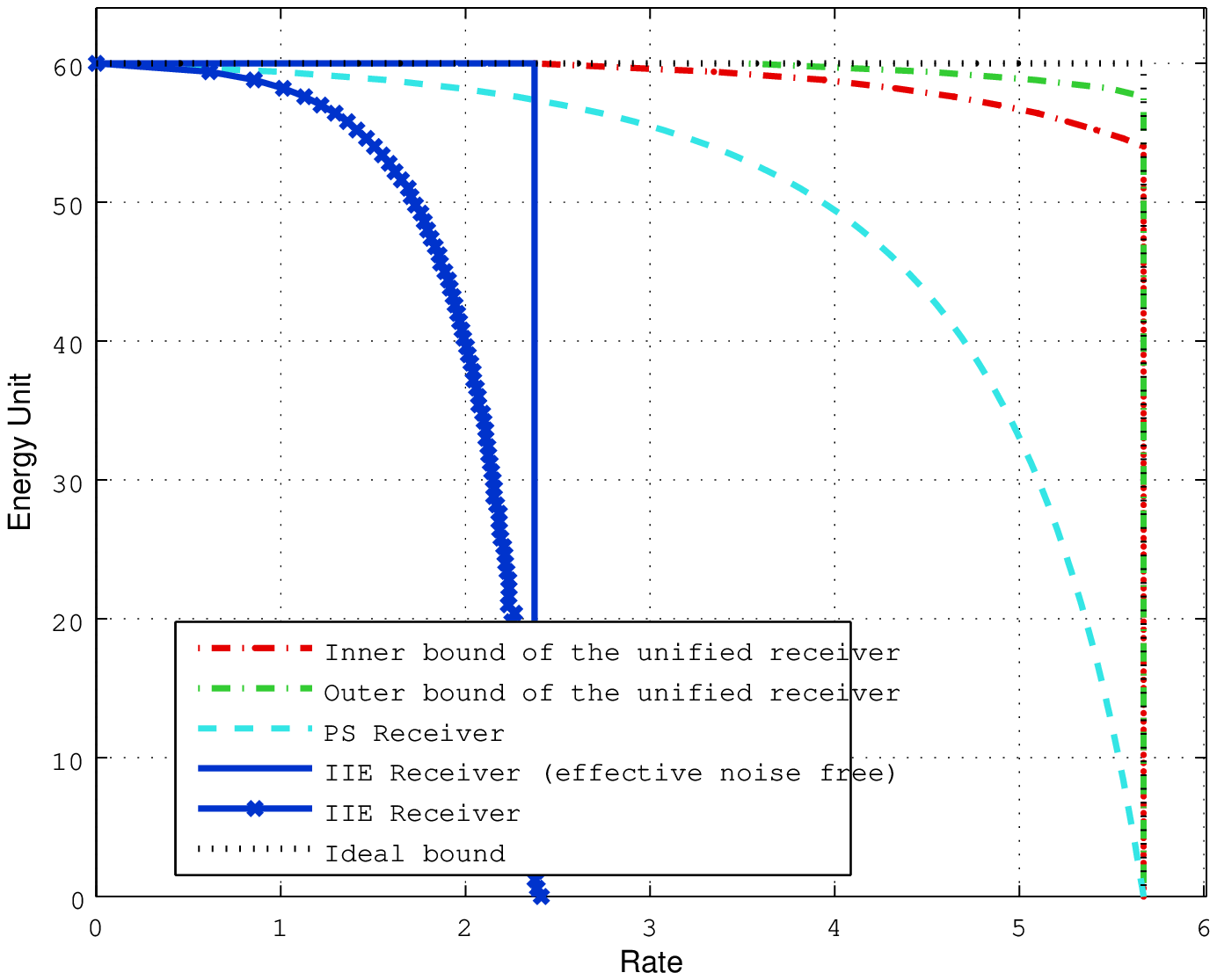}}
    \caption{Rate-energy region for the proposed receiver with $h=1$, $\sigma_{\mathrm{ant}}^2=\sigma_{\mathrm{rec}}^2=\sigma_{\mathrm{eff}}^2=1$, $\zeta=0.6$, and $P=100$.}
    \label{fg:sim1}
\end{figure}

\begin{figure}[t]
    \centerline{\includegraphics[width={0.8\columnwidth},height={0.6\columnwidth}]{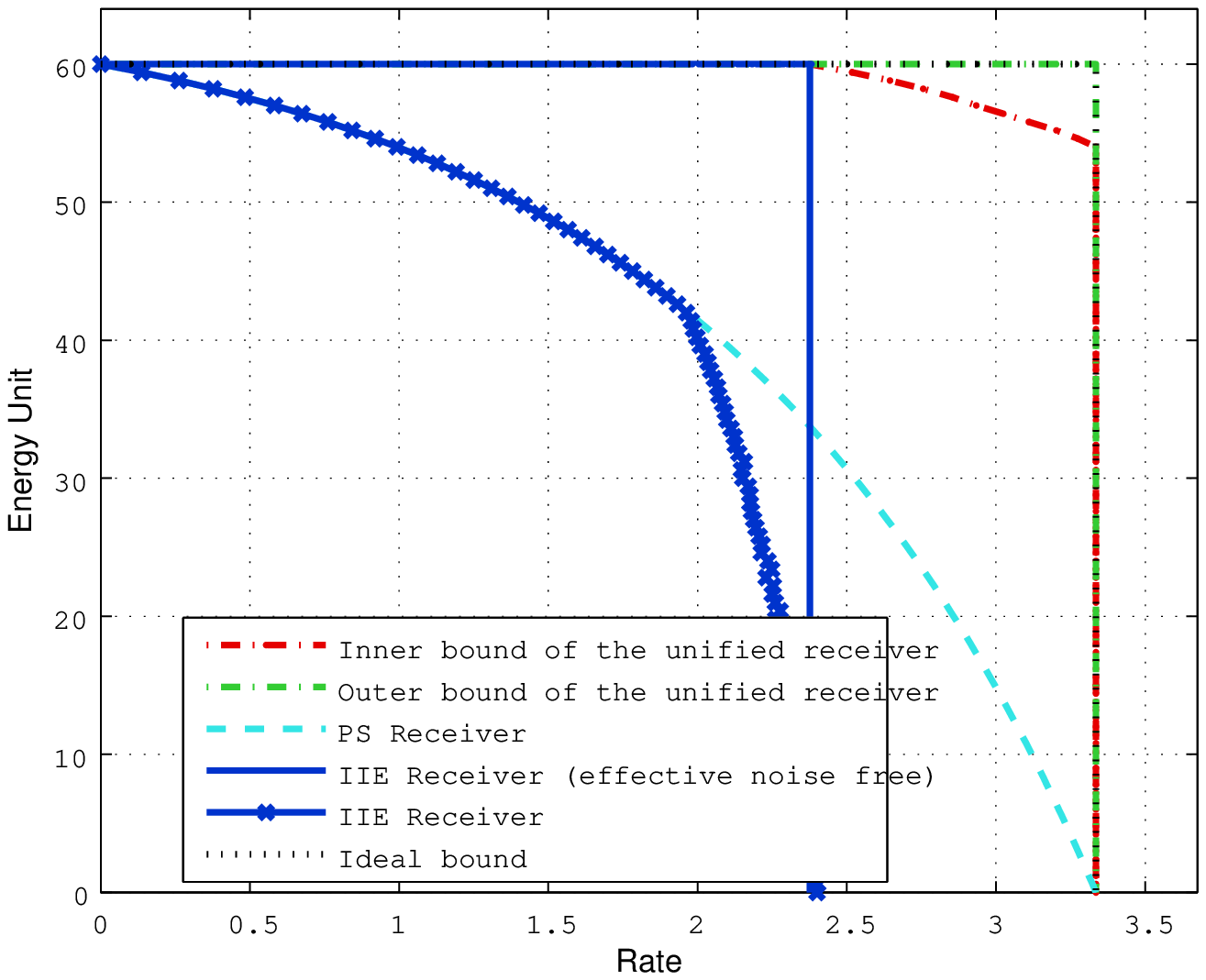}}
    \caption{Rate-energy region for the proposed receiver with $h=1$, $\sigma_{\mathrm{ant}}^2=\sigma_{\mathrm{rec}}^2=1$, $\sigma_{\mathrm{eff}}^2=10$, $\zeta=0.6$, and $P=100$.}
    \label{fg:sim1covL}
\end{figure}

\subsection{Comparisons of Rate-Energy Regions}

Figs. \ref{fg:sim1} and \ref{fg:sim1covL}  compare the proposed receiver architecture with the PS receiver and the IIE receiver in terms of rate-energy region, where $h=1$, $P=100$, $\sigma_{\mathrm{ant}}^2=\sigma_{\mathrm{rec}}^2=1$, $\zeta=0.6$, and $\sigma_{\mathrm{eff}}^2=1$ in Fig.  \ref{fg:sim1} and $\sigma_{\mathrm{eff}}^2=10$ in Fig. \ref{fg:sim1covL}.
Note that effect of ADC noise is incorporated in the effective noise. 
The label of `Ideal bound' denotes the ideal outer bound in  \eqref{eq:region_def} where energy is maximally harvested without a rate loss. The label of `Outer bound of the unified receiver' represents the upper bound on achievable rate-energy region in \eqref{eq:region} by the proposed receiver and the label of `Inner bound of the unified receiver' means the lower bound on the  achievable rate-energy region  in \eqref{eq:R_low} with the Gaussian input distribution by the proposed receiver. The rate-energy region achievable with the proposed receiver architecture certainly lies between the `Outer bound of the unified receiver' and
`Inner bound of the unified receiver' of which gap is quite small as exhibited in Figs. \ref{fg:sim1} and \ref{fg:sim1covL}. The labels of
`IIE receiver' and `PS receiver' denote the outer bounds of the rate-energy regions with the IIE receiver and the PS receiver, respectively. If $\rho=1$ in the proposed receiver, the whole received signal is rectified, so the proposed receiver becomes identical to the IIE receiver and correspondingly the harvested energy is maximized as $Q=\zeta h^2P$. If $\rho=0$ in the proposed receiver, the proposed receiver does not harvest energy and thus the achievable rate is maximized as $R=\log_2\left(1+\frac{h^2P}{\sigma_{\mathrm{ant}}^2+\sigma_{\mathrm{eff}}^2}\right)$. An arbitrary point (i.e., rate-energy tuple) on the rate-energy region with the proposed receiver architecture can be achieved by selecting an appropriate value of $\rho$ in $0<\rho<1$.  The achievable rate-energy region with the proposed architecture is very close to the ideal outer bound and remarkably larger than both the outer bounds with the IIE receiver and the PS receiver in Figs. \ref{fg:sim1} and \ref{fg:sim1covL}. The rate-energy region achievable with the proposed receiver is very close to the ideal outer bound, which indicates that the information and energy losses in SWIPT are small.
Comparing Fig. \ref{fg:sim1} with Fig. \ref{fg:sim1covL}, as the effective noise power $\sigma_{\mathrm{eff}}^2$ which accounts for quantization errors and ADC noise increases, the rate-energy region with the proposed receiver architecture is compressed along the rate axis because the achievable rate from baseband signal processing decreases as the effective noise power increases. However, the rate-energy region with the proposed receiver is still considerably larger than both upper bounds with the IIE receiver and the PS receiver and close to the ideal outer bound.

\begin{figure}[h]
    \centerline{\includegraphics[width={0.8\columnwidth},height={0.55\columnwidth}]{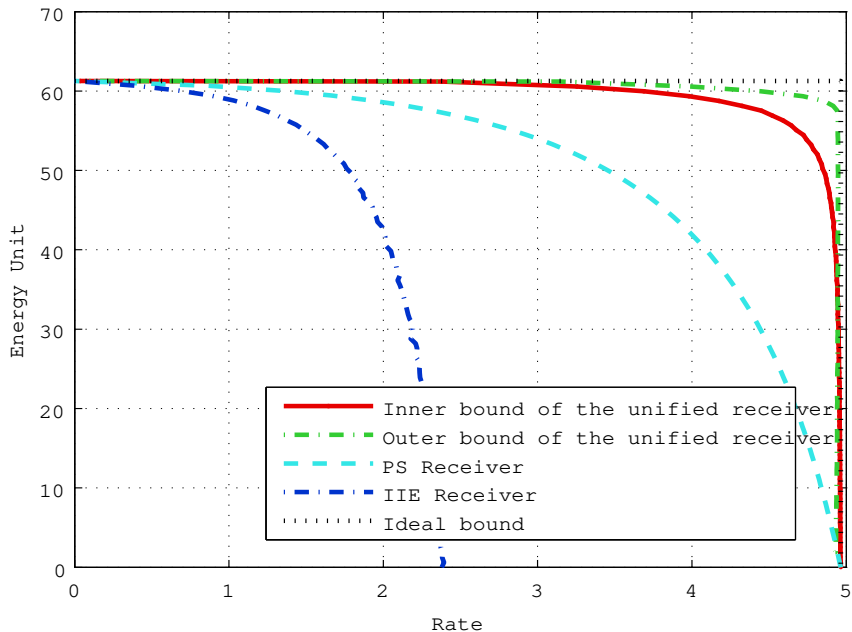}}
    \caption{Achievable ergodic rate-energy regions in a frequency non-selective fast fading channel ($h(t) \sim\mathcal{CN}(0,1) $) when $\sigma_{\mathrm{ant}}^2=\sigma_{\mathrm{rec}}^2=\sigma_{\mathrm{eff}}^2=1$, $\zeta=0.6$, and $P=100$.}
    \label{fg:simavg}
\end{figure}

To examine the effect of channel fading, we additionally consider frequency non-selective fast fading channels.
For frequency non-selective fast fading channels, the ergodic rate-energy region, that is, $\big(\mathbb{E}_h \left[ R(h(t))\right]$, $\mathbb{E}_h \left[ Q(h(t))\right]\big)$,  where $h(t)$ is the time varying  channel coefficient, is an appropriate performance metric. To verify the superiority of the proposed receiver architecture even in a frequency non-selective fast fading channel, we present the achievable ergodic rate-energy region in Fig. \ref{fg:simavg}. 
In this figure, the channel is assumed to follow a complex Gaussian channel, that is, the channel coefficient $h(t) \sim \mathcal{CN}(0,1) $, and the rate-energy regions are averaged over $10^4$ channel realizations to obtain the ergodic rate-energy region. 
As in the constant channel model, the ergodic rate-energy region of the proposed receiver is considerably larger than those of the conventional receivers. By the definition of ergodic rate-energy region, each snap shot for a channel realization corresponds to the rate-energy region in the constant channel model, so our analysis in a constant channel model builds a analytic framework to obtain the ergodic rate-energy region in time varying fading channels.

Moreover, although our analysis is based on narrow-band signals for SWIPT, our analysis can be applicable to frequency selective channels for wide-band signals for SWIPT, since orthogonal frequency division multiplexing (OFDM) can be used for wide-band signals and then each subcarrier typically experiences a frequency non-selective channel.

\section{Practical examples of the achievable rate-energy region improvement} \label{sec:Constell}

To practically account for the theoretically achievable rate-energy region, this section presents 
practical examples of the rate-energy region improvement. To this end, based on multi-dimensional Gaussian ML detection, we consider an $M$-ary multi-level modulation which leverages the proposed receiver. The proposed receiver structure is leveraged by signal constellations with multiple amplitude levels and different phases on each amplitude level. However, since taking account of all possible such constellations is impossible, we consider and optimize a structured one, multi-level circular QAM, as an example of such signal constellations. If another constellation is adopted,  the practically realized rate-energy region might vary and other constellations could yield more improved practical realization. However, for any constellation, the trend that the near-optimal rate-energy region can be achieved with the proposed receiver structure is retained.  

\begin{figure}[t]
\centering
\begin{subfigure}[][Inter-level structure]
    {\includegraphics[width={0.42\columnwidth},height={0.45\columnwidth}]{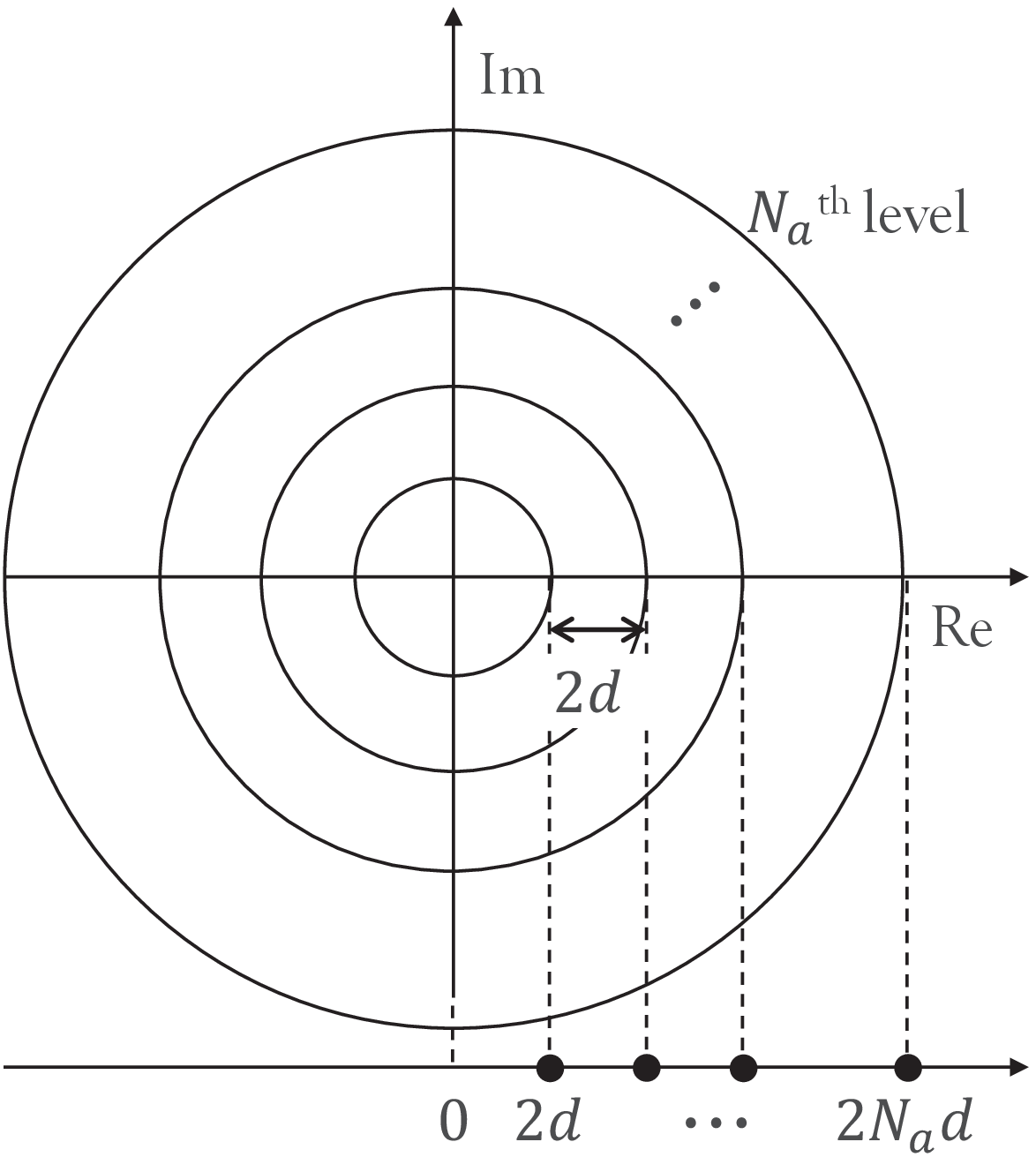}}
\end{subfigure}
\begin{subfigure}[][Constellation on the $k$th ring]
    {\includegraphics[width={0.45\columnwidth},height={0.45\columnwidth}]{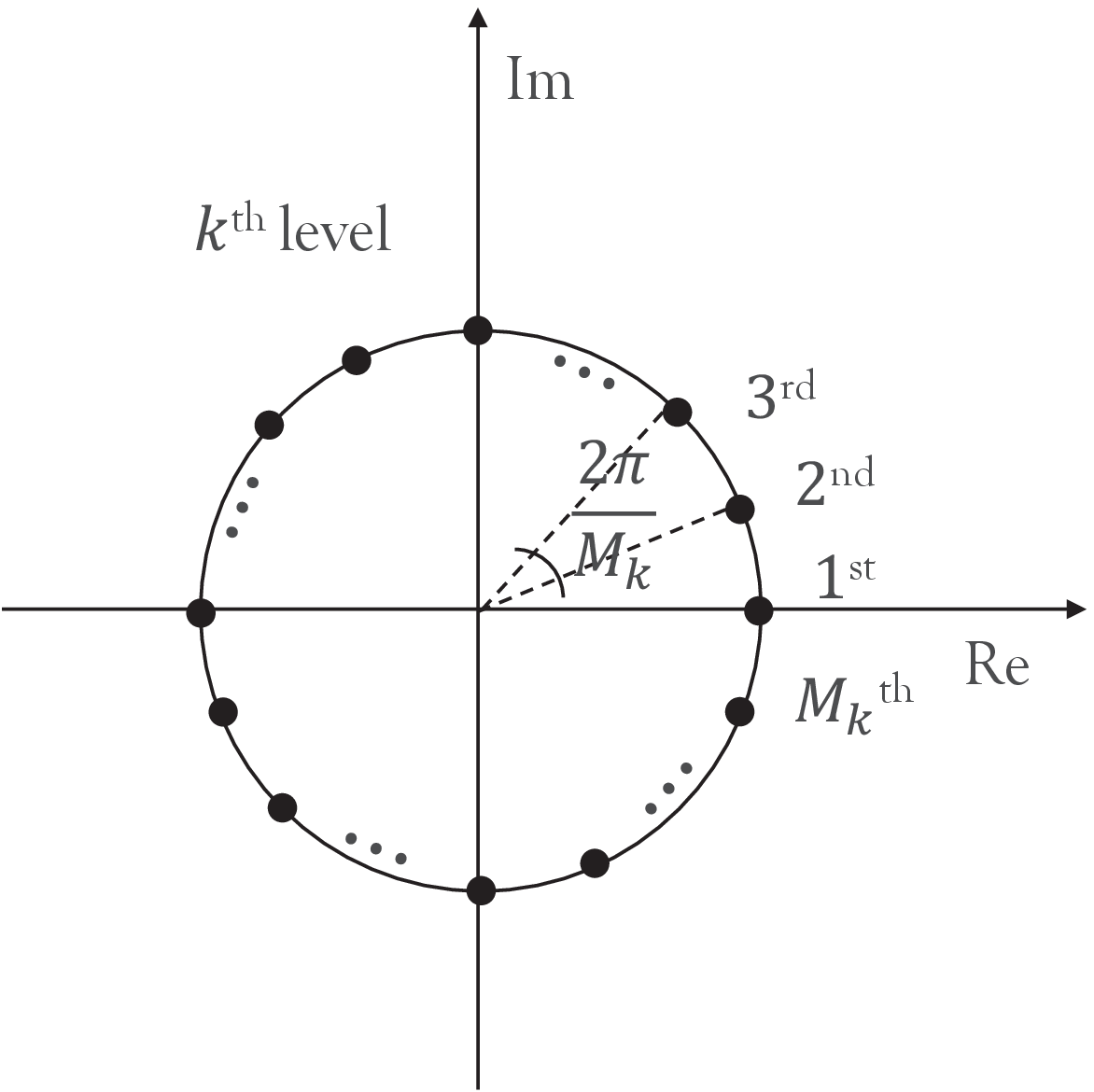}}
\end{subfigure}
    \caption{Signal constellation for the proposed $M$-ary multi-level modulation.}
    \label{fg:Constell}
\end{figure}

The constellation of the $M$-ary multi-level circular QAM  has $N_a$ amplitude levels and there are $M_k$ signal points with different phases on the ring representing
the
$k$th amplitude level as shown in Fig. \ref{fg:Constell}. In the $M$-ary multi-level circular QAM, there are total $M (=\sum_{k=1}^{N_a}M_k) $ signal points over $N_a$ amplitude levels. 
For a required amount of harvested energy $Q_{\mathrm{EH}}$, the value of $\rho$ is determined since $Q_{\mathrm{EH}}$ is given by $Q_{\mathrm{EH}}= \rho\zeta h^2P~(J)$. Then, 
signal constellation is designed by optimizing $N_a$ and $\{M_k\}$ according to the value of $\rho$ in the proposed receiver architecture.

Let $s_m$ be the modulated symbol and each symbol is generated equiprobably from $\mathcal{S}=\{s_m|m=1,\ldots, M\}$.
Then, in the propose receiver, the baseband signal as well as the power level information from the rectified signal construct a three dimensional (i.e., inphase, quadrature, and the power level from the rectified signal) sufficient statistic for demodulation as
\begin{align}
\mathbf{y}=\mathbf{H}\mathbf{s}_m+\mathbf{n}=\mathbf{u}_m+\mathbf{n}
\end{align}
where $\mathbf{H}=\mathrm{diag}\{\sqrt{P(1-\rho)}|h|,\sqrt{P(1-\rho)}|h|, P\rho|h|^2\}$ where $\mathrm{diag}\{e_1,\ldots,e_N\}$
denotes the diagonal matrix with element $e_i$ on the $i$th diagonal, 
$~\mathbf{s}_m=[\mathfrak{R}\{s_m\}, \mathfrak{I}\{s_m\}, |s_m|^2]^T$
where $\mathfrak{R}\{\cdot\}$ and $\mathfrak{I}\{\cdot\}$ are real and imaginary parts of its argument, respectively, and 
$\mathbf{n}=[n_1, n_2, n_3]$ where
$n_1=\sqrt{1-\rho}\mathfrak{R}\{z_{\mathrm{ant}}\}+\mathfrak{R}\{z_{\mathrm{eff}}\}$, $n_2=\sqrt{1-\rho}\mathfrak{I}\{z_{\mathrm{ant}}\}+\mathfrak{I}\{z_{\mathrm{eff}}\}$ and 
$n_3=\alpha_1 \mathfrak{R}\{z_{\mathrm{ant}}\}+\alpha_2 \mathfrak{I}\{z_{\mathrm{ant}}\}+z_{\mathrm{rec}}$ where $\alpha_1=2\sqrt{ P}\rho hx_a\cos\left(x_\phi + \theta\right)$ and $\alpha_2=2\sqrt{ P}\rho hx_a\sin\left(x_\phi + \theta\right)$ from (\ref{eq:output_rec}). 

Note that $\mathfrak{R}\{z_{\mathrm{ant}}\}^2$ and $\mathfrak{I}\{z_{\mathrm{ant}}\}^2$ are assumed to be negligible for analytical tractability as \cite{TC_ZZH2013} although 
$n_3=\alpha_1 \mathfrak{R}\{z_{\mathrm{ant}}\}+ \mathfrak{R}\{z_{\mathrm{ant}}\}^2+\alpha_2 \mathfrak{I}\{z_{\mathrm{ant}}\}+\mathfrak{I}\{z_{\mathrm{ant}}\}^2+z_{\mathrm{rec}}$. This assumption is well justified as follows. 
Based on 3GPP standards \cite{lte},  given the transmission bandwidth $B$=100 MHz and noise power spectral density $N_0 \approx 2\times 10^{-14}$, the variance of $\mathfrak{R}\{z_{\mathrm{ant}}\}$ and $\mathfrak{I}\{z_{\mathrm{ant}}\}$ can be formulated by $\sigma_{\mathrm{ant}}^2/2=N_0B/2\approx 10^{-6}$. The complement cumulative distribution function (CCDF) of $\left|\frac{\mathfrak{R}\{z_{\mathrm{ant}}\}^2}{\alpha_1 \mathfrak{R}\{z_{\mathrm{ant}}\}}\right|$ becomes
\begin{align}
\label{eq:probability}
\textrm{Pr}\left(\Bigg|\frac{\mathfrak{R}\{z_{\mathrm{ant}}\}^2}{\alpha_1 \mathfrak{R}\{z_{\mathrm{ant}}\}}\Bigg|\geq 0.1\right)=\textrm{Pr}\left(\Big|\mathfrak{R}\{z_{\mathrm{ant}}\}\Big|\geq 0.1\left|2\sqrt{P}\rho h\right| \right).
\end{align}
Assume $|h|=1$ as Section III. B and $\rho \approx$ 1 for enough amount of harvested energy. Let the transmitted power be $P_t$ and then $P=P_t d^{-\alpha}$ where $d$ and $\alpha$ are the distance between transmitter and receiver and the pathloss exponent, respectively. To evaluate the probability in \eqref{eq:probability}, we set
$\alpha$ to be 3 since the pathloss exponent in urban and cellular radio is from 2.7 to 3.5 and assume $d=5~\rm{(m)}$ which is considered practically appropriate for RF-based SWIPT. 
Then, for different transmit power levels, i.e., $P_t=20, 1$, and $0.5 ~\rm{(Watt)}$, $P=P_t d^{-\alpha}=0.16$, $8\times 10^{-3}, $ and $4\times 10^{-3}~\rm{(Watt)}$ for $P_t=20, 1$, and $0.5~\rm{(Watt)}$, respectively. Correspondingly, $\left|2\sqrt{P}\rho h\right|=0.8$, 0.18, and 0.13, respectively. Since 
$\textrm{Pr}\left(\Big|\mathfrak{R}\{z_{\mathrm{ant}}\}\Big|\geq 0.08  \right)$, 
$\textrm{Pr}\left(\Big|\mathfrak{R}\{z_{\mathrm{ant}}\}\Big|\geq 0.018  \right)$, and 
$\textrm{Pr}\left(\Big|\mathfrak{R}\{z_{\mathrm{ant}}\}\Big|\geq 0.013  \right)$ are approximately equal to 1, the probability of $\left|\frac{\mathfrak{R}\{z_{\mathrm{ant}}\}^2}{\alpha_1 \mathfrak{R}\{z_{\mathrm{ant}}\}}\right|\leq 0.1$ is almost one with high probability.   Therefore, we can justify the assumption of $\mathfrak{R}\{z_{\mathrm{ant}}\}^2+\mathfrak{I}\{z_{\mathrm{ant}}\}^2$  and $n_3$ is simplified as
$\alpha_1 \mathfrak{R}\{z_{\mathrm{ant}}\}+\alpha_2 \mathfrak{I}\{z_{\mathrm{ant}}\}+z_{\mathrm{rec}}$. In addition, in view of average signal power, the ratio between noise power and squared noise power scales $10^{-6}$. Therefore, $\mathfrak{R}\{z_{\mathrm{ant}}\}^2$ and $\mathfrak{I}\{z_{\mathrm{ant}}\}^2$ can be reasonably assumed to be negligible for analytical tractability.

It is known that Maximal Likelihood (ML) is the optimal detection if symbols are
generated equiprobably and channel state information at receiver (CSIR) is available. 
Since all elements of $\mathbf{n}$ include $\mathfrak{R}\{z_{\mathrm{ant}}\}$ and $\mathfrak{I}\{z_{\mathrm{ant}}\}$, the noise vector $\mathbf{n}$ is a correlated Gaussian noise vector. 
After whitening the correlated noise vector based on its covariance matrix given by $\mathbf{\Sigma_n}=\mathbf{\Lambda_n}\mathbf{\Lambda_n}^T$, the ML decision rule is formulated as 
\begin{align}
\max_{s_m \in \mathcal{S}} \ln f(\mathbf{y}|s_m)&=\min_{s_m \in \mathcal{S}}
(\mathbf{y}-\mathbf{u}_m)^T\mathbf{\Sigma_n}^{-1}(\mathbf{y}-\mathbf{u}_m)\\
&=\min_{s_m \in \mathcal{S}}
||\mathbf{\Lambda_n}^{-1}(\mathbf{y}-\mathbf{u}_m)||^2,
\end{align}
where $f(\mathbf{y}|s_m)$  is the likelihood function given by a conditional probability density function (PDF) $\sim\mathcal{N}(\mathbf{u}_m, \mathbf{\Sigma_n})$;
$\mathbf{n}$ is a jointly Gaussian random vector $\sim\mathcal{N}(\mathbf{0}, \mathbf{\Sigma_n})$  where $\mathbf{\Sigma_n}$ is its covariance matrix given by
\begin{align}
\mathbf{\Sigma_n} &\overset{(f)}= \left[
                  \begin{array}{ccc}
                   E[n_1^2] &E[n_1n_2]&E[n_1n_3] \\
                    E[n_1n_2]  &  E[n_2^2]&E[n_2n_3] \\
                   E[n_1n_3]  & E[n_3n_2] &E[n_3^2]
                  \end{array}
                \right]\\
               &= \left[
                  \begin{array}{ccc}
                   \{(1-\rho)\sigma_{\mathrm{ant}}^2+\sigma_{\mathrm{eff}}^2\}/2 &0&\alpha_1\sqrt{1-\rho}\sigma_{\mathrm{ant}}^2/2 \\
                   0  &  \{(1-\rho)\sigma_{\mathrm{ant}}^2+\sigma_{\mathrm{eff}}^2\}/2&\alpha_2\sqrt{1-\rho}\sigma_{\mathrm{ant}}^2/2 \\
                  \alpha_1\sqrt{1-\rho}\sigma_{\mathrm{ant}}^2/2  & \alpha_2\sqrt{1-\rho}\sigma_{\mathrm{ant}}^2/2 &\{(\alpha_1^2+\alpha_2^2)\sigma_{\mathrm{ant}}^2+2\sigma_{\mathrm{rec}}^2\}/2
                  \end{array}
                \right]\\
                & =\mathbf{\Lambda_n}\mathbf{\Lambda_n}^T
\end{align}
where $(f)$ holds from $\mathrm{E}[\mathbf{n}]=\mathbf{0}$.
Then, the pairwise error probability (PEP) based on the ML detection  that $s_j$ is detected when $s_i$ was transmitted under CSIR is given by
\begin{align}
\textrm{Pr}(s_i \rightarrow s_j |\mathbf{H})=\textrm{Pr}(||\mathbf{\Lambda_n}^{-1}(\mathbf{y}-\mathbf{u}_i)||^2>||\mathbf{\Lambda_n}^{-1}(\mathbf{y}-\mathbf{u}_j)||^2 |\mathbf{H})=Q\left(\frac{1}{2}||\mathbf{\Lambda_n}^{-1}(\mathbf{u}_i-\mathbf{u}_j)||\right),
\end{align}
where $\forall i \neq j$.

Based on the multi-dimensional ML detection, the $M$-ary multi-level circular QAM is designed to maximize the data rate with a given
transmit power $P$, an energy portion of the received signal $\rho$, and a target symbol error rate $P_e^{\left(target\right)}$. That is, the design parameters, $N_a$ and $\{M_k\}$, and correspondingly $M =\sum_{k=1}^{N_a} M_k$, are determined by solving the following optimization problem:
\begin{align} \label{eq:problem}
(\mathbf{P1}):~\max_{N_{a},\left\{M_k\right\}} \quad & \log_2{M} \\
\label{eq:constraint1}
\mathrm{such~that} \quad\quad\quad\quad & Q\left(\frac{1}{2}||\mathbf{\Lambda_n}^{-1}(\mathbf{u}_i-\mathbf{u}_j)||\right)\leq P_e^{\left(target\right)},\forall i \neq j , \\
\label{eq:constraint2}
& M=\sum_{k=1}^{N_a}M_k, \\
\label{eq:constraint3}
& \frac{1}{M}\sum_{k=1}^{N_a}M_k \left(2kd\right)^2 \leq P.
\end{align}

Note that if $\rho=1$, the optimization problem $\mathbf{P1}$ reduces to design of conventional PAM. If $\rho=0$, the optimization problem $\mathbf{P1}$ refers to design of the conventional circular QAM without help of amplitude information from 
the rectified signal.

Since $M$, $N_a$, and $M_k$ are integers, $\mathbf{P1}$ is an integer programming problem that is known to barely have a closed form solution. Moreover,  $Q\left(\frac{1}{2}||\mathbf{\Lambda_n}^{-1}(\mathbf{u}_i-\mathbf{u}_j)||\right)\leq P_e^{\left(target\right)}$ is a non-convex function and thus we have to rely on numerical methods to solve . However, fortunately, $M$ can be upper-bounded and search complexity for a bounded integer is not so high; practically feasible $\log_2 M$ is about 10. To reduce the search complexity further, we can consider the $M$-ary multi-level modulation with the same number of constellation points on each ring, i.e., $M_1 =\ldots = M_{N_a} = \frac{M}{N_a}$. It is also assumed that each ring has the same phase offset for the signal points on each ring. 
Then, we determine $N_a$ and $M$ by solving the following problem:
\begin{align} 
(\mathbf{P2}):~\max_{N_{a}} \quad & \log_2{M} \\
\mathrm{such~that} \quad\quad\quad\quad & Q\left(\frac{1}{2}||\mathbf{\Lambda_n}^{-1}(\mathbf{u}_i-\mathbf{u}_j)||\right)\leq P_e^{\left(target\right)},\forall i \neq j , \\
& \frac{1}{N_a}\sum_{k=1}^{N_a} \left(2kd\right)^2 \leq P.
\end{align}
Note that the considered M-ary multi-level circular QAM is not optimal  but for demonstrating the rate-energy region improvement with practical modulation.

\begin{figure}[t]
    \centerline{\includegraphics[width={1\columnwidth},height={0.75\columnwidth}]{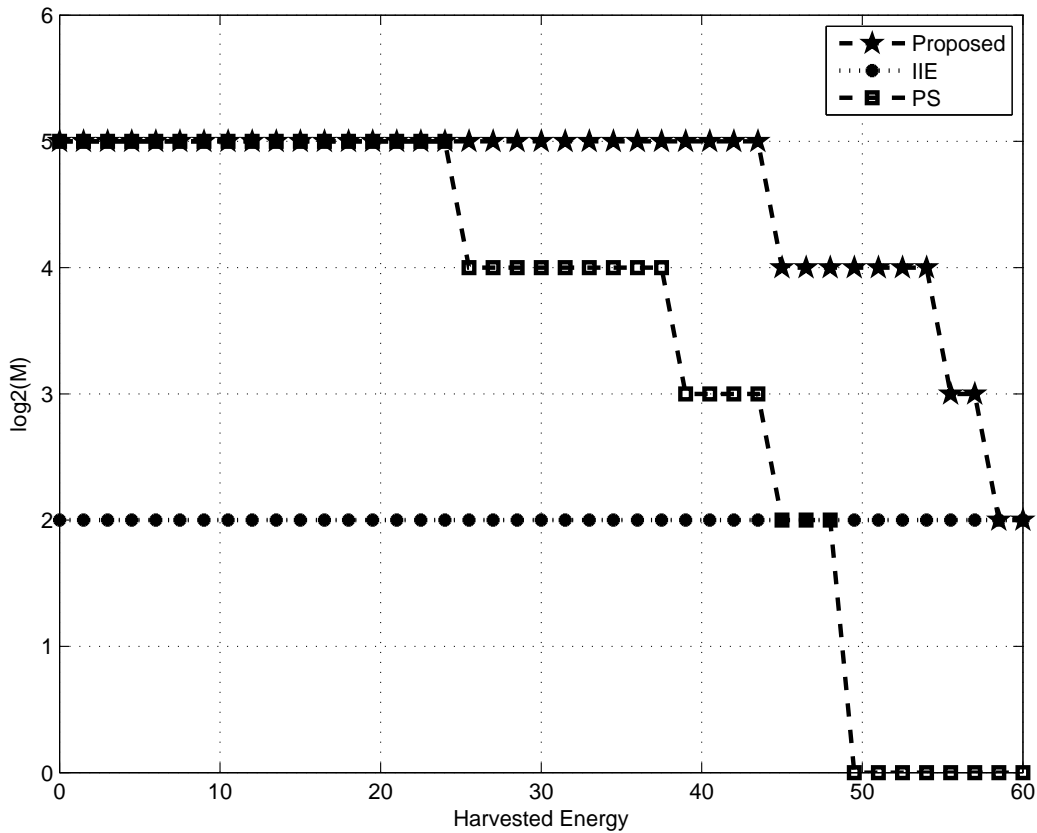}}
    \caption{Maximum modulation order $\log_2 M$ for the proposed receiver  and the referential receivers according to varying energy portion of the received signal $\rho$ when $\frac{E_b}{N_0}=20$ dB and $P_e^{\left(target\right)}=10^{-3}$ when $\zeta=0.6$ and $h=1$.The  amount of harvested energy, $Q_{\mathrm{EH}}=\rho\zeta h^2P~(J)=60\rho$ .}
    \label{fg:simM}
\end{figure}

The maximum modulation order $\log_2 M$ is plotted versus the required amount of harvested energy, $Q_{\mathrm{EH}}= \rho\zeta h^2P~(J)=60\rho$ when $\zeta=0.6$, $h=1$ and $P=100$ in Fig. \ref{fg:simM}, after numerically solving the optimization problem $\mathbf{P2}$ with the target symbol error probability of $P_e^{\left(target\right)}=10^{-3}$. 
The label of `Proposed' denotes the proposed unified SWIPT receiver structure exploiting the optimized $M$-ary multi-level circular QAM based on the three-dimensional ML detection.  The labels of `IIE' and `PS' denote the IIE and PS receivers, respectively. Note that the IIE receiver exploits PAM
modulation/demodulation since the rectified signal is split. For the PS receiver, the $M$-ary multi-level modulation optimized based on the optimization problem $\mathbf{P2}$ for the PS receiver is adopted. The proposed scheme achieves $M=32$ when $ 0 \leq Q_{\mathrm{EH}} \leq 46.5)$ (\emph{i.e.,} $0\leq \rho \leq 0.775$). Although the achievable $M$ decreases with $\rho$ only beyond $\rho=0.775$, the proposed scheme outperforms the other two referential schemes for all $Q_{\mathrm{EH}}$.  On the other hand,  'IIE'  achieves higher modulation order $\log_2 M$ than `PS' if the amount of energy  to be harvested is high, \emph{i.e.,}  $50< Q_{\mathrm{EH}} \leq 60$.

\begin{figure}[t]
    \centerline{\includegraphics[width={1\columnwidth},height={0.75\columnwidth}]{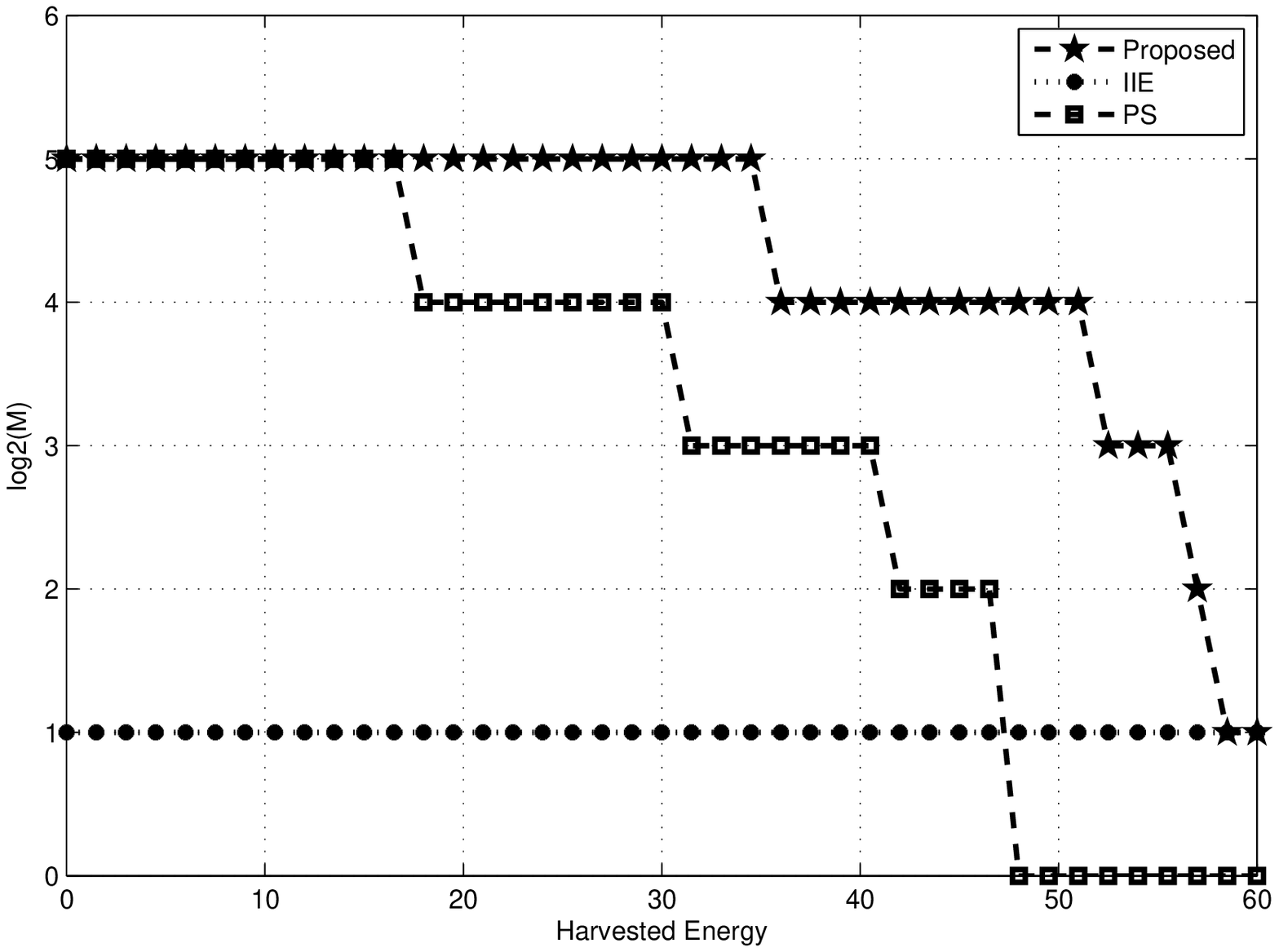}}
    \caption{Maximum modulation order $\log_2 M$ for the proposed receiver and the referential receivers according to varying energy portion of the received signal $\rho$ when $\frac{E_b}{N_0}=20$ dB and $P_e^{\left(target\right)}=10^{-4}$ when $\zeta=0.6$ and $h=1$. The amount of harvested energy, $Q_{\mathrm{EH}}=\rho\zeta h^2P~(J)=60\rho$ .}
    \label{fg:simM_low}
\end{figure}
\begin{table}[h]
 \caption{Optimal modulation constellation ($\log_2M^*$, $\log_2N_a^*$) according to $\rho ~(=Q_{\mathrm{EH}}/\zeta h^2P=Q_{\mathrm{EH}}/60)$ for target PEP $10^{-3}$ }
    \centering
    \label{Na1}
      \begin{tabular}{|c|c|c|c|c|c|c|c|c|c|}
       \hline
        $\rho\!\!$ &  $\![0\!\!:\!\!\frac{1}{10}]\!\!$ & $\![\frac{1}{10}\!\!:\!\!\frac{2}{5}]\!\!$ &$\![\frac{2}{5}\!\!:\!\!\frac{5}{8}]\!\!$&$\![\frac{5}{8}\!\!:\!\!\frac{29}{40}]\!\!$ &$\![\frac{29}{40}\!\!:\!\!\frac{4}{5}]\!\!$ &$\![\frac{4}{5}\!\!:\!\!\frac{9}{10}]\!\!$ &$\![\frac{9}{10}\!\!:\!\!\frac{37}{40}]\!\!$ &$\![\frac{37}{40}\!\!:\!\!\frac{19}{20}]\!\!$ & $\![\frac{19}{20}\!\!:\!\!1]\!\!$\\
        \hline
        `Proposed' &(5,2)  & (5,4) &(5,4) & (5,4)&(4,3) &(4,3) &(4,3) &(3,1) &(2,1)\\
        \hline
       `PS' & (5,4) & (5,4) & (4,2)& (3,1)& (2,1)& (0,0)& (0,0)& (0,0)&(0,0) \\
        \hline
    \end{tabular}
\end{table}
 \begin{table}[h]
 \caption{Optimal modulation constellation ($\log_2 M^*$, $\log_2N_a^*$) according to $\rho ~(=Q_{\mathrm{EH}}/\zeta h^2P=Q_{\mathrm{EH}}/60)$ for target PEP $10^{-4}$}
    \centering
    \label{Na2}
         \begin{tabular}{|c|c|c|c|c|c|c|c|c|c|c|}
       \hline
         $\rho\!\!$ &  $\![0\!\!:\!\!\frac{1}{10}]\!\!$ & $\![\frac{1}{10}\!\!:\!\!\frac{11}{40}]\!\!$ & $\![\frac{11}{40}\!\!:\!\!\frac{1}{2}]\!\!$ &$\![\frac{1}{2}\!\!:\!\!\frac{23}{40}]\!\!$ &$\![\frac{23}{40}\!\!:\!\!\frac{27}{40}]\!\!$ &$\![\frac{27}{40}\!\!:\!\!\frac{31}{40}]\!\!$ &$\![\frac{31}{40}\!\!:\!\!\frac{17}{20}]\!\!$ &$\![\frac{17}{20}\!\!:\!\!\frac{37}{40}]\!\!$ & $\![\frac{37}{40}\!\!:\!\!\frac{19}{20}]\!\!$&$\![\frac{19}{20}\!\!:\!\!1]\!\!$\\
        \hline
        `Proposed' &(5,2)  & (5,4) &(5,4) & (5,4)&(4,3) &(4,3) &(4,3) &(3,2) &(2,1)&(1,1)\\
        \hline
       `PS' & (5,4) & (5,4) & (4,2)& (3,1)& (3,1)& (2,1)& (0,0)& (0,0)&(0,0)& (0,0)\\
        \hline
    \end{tabular}
\end{table}
  
Fig. \ref{fg:simM_low} exhibits the maximum modulation order $\log_2 M$ as a function of $\rho$ when $P_e^{\left(target\right)}=10^{-4}$. Except for the target symbol error probability, this figure has the same settings as Fig. \ref{fg:simM}.  Fig. \ref{fg:simM_low} shows degraded performance compared to  Fig. \ref{fg:simM} due to the tighter target symbol error probability, but the overall trend is the same as Fig.\ref{fg:simM}.
Comparing Figs. \ref{fg:simM} and \ref{fg:simM_low} with Figs. \ref{fg:sim1} and \ref{fg:sim1covL}, the inverses of curves in Figs. \ref{fg:simM} and \ref{fg:simM_low} are roughly similar to Figs. \ref{fg:sim1} and \ref{fg:sim1covL}.
 That is, the maximum size $\log_2 M$ satisfying target PEP according to the amount of harvested energy practically 
 accounts for the information theoretic rate-energy tradeoff region. Consequently, Figs. \ref{fg:simM} and \ref{fg:simM_low} reveal the rate-energy tradeoff from a practical viewpoint. 

To see the optimal modulation constellation according to $\rho$, Tables \ref{Na1} and \ref{Na2} present optimal $\log_2 M$ and $\log_2 N_a$ together after solving the optimization problem $\mathbf{P2}$ for target PEPs $10^{-3}$ and $10^{-4}$  when $\zeta=0.6$, $h=1$ and $P=100$. If there are different values of $N_a$ yielding the maximum $M$ while satisfying the constraints, the one achieving the smallest PEP is selected as the optimal value of $N_a$. 
 Let  $\log_2 M^*(\rho, P_e^{\left(target\right)})$ and $\log_2 N_a^*(\rho, P_e^{\left(target\right)})$ denote the maximum modulation order and the optimal bits allocated to amplitude information for given $\rho$ and target PEP, respectively. That is, the optimal number of rings is $N_a^*(\rho, P_e^{\left(target\right)})$. The optimized $M$-ary multi-level  circular QAM consists of $N_a^*(\rho, P_e^{\left(target\right)})$ rings with different amplitudes and $M^*(\rho, P_e^{\left(target\right)})/N_a^*(\rho, P_e^{\left(target\right)})$ constellation points are  placed on each ring. In the proposed scheme, optimal $N_a$ decreases as the required amount of energy to be harvested increases in general.

\section{Conclusion} \label{sec:Con}
In this paper, we proposed a unified receiver architecture for simultaneous wireless information and power transfer, and derived tight upper and lower bounds on the rate-energy region achieved with the proposed receiver architecture. It was proved that the 
the achievable rate-energy region is considerably expanded over those of conventional schemes and becomes close to the ideal upper bound.  In the proposed receiver architecture, the energy required for information decoding at the decoding circuit can be minimized because the amplitude information from the energy harvesting circuit is also exploited in information decoding. Consequently, the fundamental tradeoff in SWIPT is nearly overcome and thus the near optimal rate-energy region is achievable. 
To practically account for the theoretically achievable rate-energy region, we also presented practical examples of the rate-energy region improvement using an $M$-ary multi-level circular QAM based on the multi-dimensional Gaussian ML detection.



\begin{thebibliography}{10}
\bibitem{ISIT_V2008} 
    L. R. Varshney, ``Transporting information and energy simultaneously,'' in \emph{Proc. IEEE Int. Symp. Inf. Theory (ISIT)}, Toronto, Canada, July 6-11, 2008, pp. 1612-1616.
\bibitem{ISIT_GS2010} 
    P. Grover and A. Sahai, ``Shannon meets Tesla: wireless information and power transfer,'' in \emph{Proc. IEEE Int. Symp. Inf. Theory (ISIT)}, Austin, Texas, USA, June 13-18, 2010, pp. 2363-2367.

\bibitem{CM_BHZ2015} 
    S. Bi, C. K. Ho, and R. Zhang, ``Wireless powered communication: opportunities and challenges,'' \emph{IEEE Commun. Mag.}, vol. 53, no. 4, pp. 117-125, Aug. 2015.


\bibitem{TWC_LZC2013} 
    L. Liu, R. Zhang, and K.-C. Chua, ``Wireless information transfer with opportunistic energy harvesting,'' \emph{IEEE Trans. Wireless Commun.}, vol. 12, no. 1, pp. 288-300, Jan. 2013.
\bibitem{TC_LZC2013} 
    L. Liu, R. Zhang, and K.-C. Chua, ``Wireless information and power transfer: A dynamic power splitting approach,'' \emph{IEEE Trans. Commun.}, vol. 61, no. 9, pp. 3990-4001, Sep. 2013.
\bibitem{TC_ZZH2013} 
    X. Zhou, R. Zhang, and C. K. Ho, ``Wireless information and power transfer: architecture design and rate-energy tradeoff,'' \emph{IEEE Trans. Commun.}, vol. 61, no. 11, pp. 4754-4767, Nov. 2013.

\bibitem{TWC_ZH2013} 
R. Zhang and C. K. Ho, ``MIMO broadcasting for simultanenous wireless information and power transfer,'' \emph{IEEE Trans. Wireless Commun.}, vol. 12, no. 5, pp. 1989-2001, May 2013.

\bibitem{JSEC_ZYH2015} 
    R. Zhang, L.-L. Yang, L. Hanzo, ``Energy pattern aided simultaneous wireless information and power transfer,'' \emph{IEEE J. Sel. Areas Commun.}, vol. 33, no. 8, pp. 1492-1504, Aug. 2015.





\bibitem{TWC_SM2010} 
    V. Sharma, U. Mukherji, V. Joseph, and S. Gupta, ``Optimal energy management policies for energy harvesting sensor nodes,'' \emph{IEEE Trans. Wireless Commun.}, vol. 9, no. 4, pp. 1326-1336, Apr. 2010.
\bibitem{TC_HO2012}
C. K. Ho and R. Zhang, ``Optimal energy allocation for wireless communications with energy harvesting constraints,'' IEEE Transactions on Signal Processing, vol. 60, no. 9, pp. 4808-4818, Sep. 2012.
    
\bibitem{TC_YU2012} 
    J. Yang and S. Ulukus, ``Optimal packet scheduling in an energy harvesting communication system,'' \emph{IEEE Trans. Commun.}, vol. 60, no. 1, pp. 220-230, Jan. 2012.
\bibitem{TWC_TY2012} 
    K. Tutuncuoglu and A. Yener, ``Optimum transmission policies for battery limited energy harvesting nodes,'' \emph{IEEE Trans. Wireless Commun.}, vol. 11, no. 3, pp. 1180-1189, Mar. 2012.
    \bibitem{TIT_OU2012} 
    O. Ozel and S. Ulukus, ``Achieving AWGN capacity under stochastic energy harvesting,'' \emph{IEEE Trans. Inf. Theory}, vol. 58, no. 10, pp. 6471-6483, Oct. 2012.
\bibitem{JCAN_YS2012} 
    J. Yang and S. Ulukus, ``Optimal packet scheduling in a multiple access channel with energy harvesting transmitters,'' \emph{IEEE J. Commun. and Network}, vol. 14, no. 2, pp. 140-150, Apr. 2012.
\bibitem{TC_OY2012} 
    O. Ozel, J. Yang, S. Ulukus,``Optimal broadcast scheduling for an energy harvesting rechargeable transmitter with a finite capacity battery'' \emph{IEEE Trans. Wireless Commun.}, vol. 11, no. 6, pp. 2193-2203, June 2012.
\bibitem{JSEC_HZ2013} 
    C. Huang, R. Zhang, and S. Cui,``Throughput maximization for the Gaussian relay channel with energy harvesting constraints'' \emph{IEEE J. Sel. Areas Commun.}, vol. 31, no. 8, pp. 1469-1479, Aug. 2013.
\bibitem{TC_SCK2015} 
    D. K. Shin, W. Choi, and D. Kim, ``The two-user Gaussian interference channel with energy harvesting transmitters: energy cooperation and achievable rate region,'' \emph{IEEE Trans. Commun.}, vol. 63, no. 11, pp. 4551-4564, Nov. 2015.
\bibitem{Arxiv_L2011} 
    G. A. Covic and J. T. Boys, ``Inductive power transfer,'' \emph{Proc. IEEE}, vol. 101, no. 6, pp. 1276-1289, May 2013.
\bibitem{Science_K2007} 
    A. Kurs, A. Karalis, R. Moffatt, J. D. Joannopoulos, P. Fisher, and M. Solja$\mathrm{\check{c}}$i$\mathrm{\acute{c}}$, ``Wireless power transfer via strongly coupled magnetic resonances,'' \emph{Science}, vol. 317, no. 5834, pp. 83–86, June 2007.
\bibitem{ECCE_M2010} 
    J. O. Mur-Miranda, G. Fanti, Y. Feng, K. Omanakuttan, R. Ongie, A. Setjoadi, and N. Sharpe, ``Wireless power transfer using weakly coupled magnetostatic resonators,'' in \emph{Proc. IEEE Energy Convers. Congr. Expo. (ECCE)}, Atlanta, GA, USA, Sep. 12-16, 2010, pp. 4179-4186.

\bibitem{WPCN1}
    H. Ju and R. Zhang, ``Throughput maximization in wireless powered communication networks,'' \emph{IEEE Trans. Wireless Commun.}, vol. 13, no. 1, Jan. 2014.
\bibitem{WPCN2}
Q. Wu, M. Tao, D. W. K. Ng, W. Chen, and R. Schober, ``Energy-efficient resource allocation for wireless powered communication networks,'' \emph{IEEE Trans. Wireless Commun.}, vol. 15, no. 3, pp. 2312-2327, Mar. 2016.
\bibitem{TC_GOYU2013} 
    B. Gurakan, O. Ozel, J. Yang, and S. Ulukus, ``Energy cooperation in energy harvesting communications,'' \emph{IEEE Trans. Commun.}, vol. 61, no. 12, pp. 4884-4898, Dec. 2013.
\bibitem{WPCN3}
H. Ju and R. Zhang, ``User cooperation in wireless powered communication
networks,'' \emph{Proc. IEEE Global Commun. Conf.}, Austin, TX, USA, 2014,
pp. 1430–1435.
\bibitem{WPCN6} 
H. Chen, Y. Li, J. L. Rebelatto, B. F. Uchoa-Filho, and B. Vucetic,
``Harvest-then-cooperate: Wireless-powered cooperative communications,''
\emph{IEEE Trans. Signal Process.}, vol. 63, no. 7, pp. 1700–1711, Apr.
2015.
\bibitem{WPCN4} 
H. Ju and R. Zhang, ``Optimal resource allocation in full-duplex wireless powered
communication network,'' \emph{IEEE Trans. Commun.}, vol. 62,
no. 10, pp. 3528–3540, Oct. 2014.
\bibitem{WPCN5}
X. Kang, C. K. Ho, and S. Sun, ``Full-duplex wireless powered communication
network with energy causality,'' \emph{IEEE Trans. Wireless Commun.},
vol. 14, no. 10, pp. 5539–5551, Oct. 2015.
\bibitem{WPCN8}
J. Zhang, C. Yuen, and C.-K. Wen, ``Large-system analysis of ergodic
sum-rate in wireless-powered MIMO communication network,''
\emph{Proc. 11th Annu. IEEE Int. Conf. SECON Workshops}, Singapore,
Jun./Jul. 2014, pp. 57–61.
\bibitem{WPCN9} 
G.-M. Yang, C.-C. Ho, R. Zhang, and Y. Guan, ``Throughput optimization
for massive MIMO systems powered by wireless energy transfer,'' \emph{IEEE
J. Sel. Areas Commun.}, vol. 33, no. 8, pp. 1640–1650, Aug. 2015.
\bibitem{WPCN11} 
S. Lee and R. Zhang, ``Cognitive wireless powered network:
Spectrum sharing models and throughput maximization,'' \emph{arXiv preprint
arXiv:1506.05925}, 2015.
\bibitem{varshney}
    L. R. Varshney, ``On energy/information cross-layer architectures,'' in \emph{Proc. IEEE Int. Symp. Inf. Theory (ISIT)}, Cambridge, MA, USA, July 1-6, 2012, pp. 1356-1360.
    

\bibitem{TIT_LMW2009} 
    A. Lapidoth, S. M. Moser, and M. A. Wigger, ``On the capacity of free space optical intensity channels,'' \emph{IEEE Trans. Inf. Theory}, vol. 55, no. 10, pp. 4449-4461, Oct. 2009.
\bibitem{TIT_KS2004} 
    M. Katz and S. Shamai, ``On the capacity-achieving distribution of the discrete-time noncoherent and partially coherent AWGN channels,'' \emph{IEEE Trans. Inf. Theory}, vol. 50, no. 10, pp. 2257-2270, Oct. 2004.
\bibitem{C_L2002} 
    A. Lapidoth, ``Capacity bounds via duality: A phase noise example,'' in \emph{Proc. 2nd Asian-European Workshop on Information Theory}, Breisach, Germany, June 26-29, 2002, pp. 58–61.
  \bibitem{lte}  
  3GPP TS 36.213 V12.11.0 (2016-09) Evolved Universal Terrestrial Radio Access (E-UTRA); Physical layer procedures.




\end{thebibliography}
\end{document}